\documentclass[fleqn,usenatbib]{mnras}

\usepackage{newtxtext,newtxmath}
\usepackage{xcolor}

\usepackage[T1]{fontenc}

\DeclareRobustCommand{\VAN}[3]{#2}
\let\VANthebibliography\thebibliography
\def\thebibliography{\DeclareRobustCommand{\VAN}[3]{##3}\VANthebibliography}

\usepackage{graphicx}	
\usepackage{amsmath}	
\usepackage{orcidlink}

\title[ALMA resolves the binary HBC 494]{Resolving the Binary Components of the Outbursting Protostar HBC 494 with ALMA}

\author[P. H. Nogueira et al.]{Pedro Henrique Nogueira$^{1,2}$\thanks{E-mail: pedro.soares@mail.udp.cl}\orcidlink{0000-0001-8450-3606},
Alice Zurlo$^{1,2}$\orcidlink{0000-0002-5903-8316}, Sebastián P\'erez$^{2,3,4}$\orcidlink{0000-0003-2953-755X}, Camilo González-Ruilova$^{1,2,5}$\orcidlink{0000-0003-4907-189X},  \newauthor Lucas A. Cieza$^{1,2}$\orcidlink{0000-0002-2828-1153}, 
Antonio Hales$^{6,7}$\orcidlink{0000-0001-5073-2849}, Trisha Bhowmik$^{1,2,3}$\orcidlink{0000-0002-4314-9070}, Dary A. Ruíz-Rodríguez$^{7}$\orcidlink{0000-0003-3573-8163}, \newauthor David A. Principe$^{8}$\orcidlink{0000-0002-7939-377X}, Gregory J. Herczeg$^{9}$\orcidlink{0000-0002-7154-6065}, Jonathan P. Williams$^{10}$\orcidlink{0000-0001-5058-695X}, Jorge Cuadra$^{11,13}$\orcidlink{0000-0003-1965-3346}, \newauthor
Matías Montesinos$^{12,13}$\orcidlink{0000-0001-9789-5098}, Nicolás Cuello$^{14}$\orcidlink{0000-0003-3713-8073}, Prachi Chavan$^{1,2}$\orcidlink{0000-0003-2406-0684}, Simon Casassus$^{2,15,16,17}$\orcidlink{0000-0002-0433-9840}, \newauthor Zhaohuan Zhu$^{18,19}$\orcidlink{0000-0003-3616-6822}, Felipe G. Goicovic$^{20}$\orcidlink{0000-0002-1429-4148}
\\
\\
$^{1}$Instituto de Estudios Astrof\'isicos, Facultad de Ingenier\'ia y Ciencias, Universidad Diego Portales, Av. Ej\'ercito Libertador 441, Santiago, Chile \\
$^{2}$Millennium Nucleus on Young Exoplanets and their Moons (YEMS) \\
$^{3}$Departamento de Física, Universidad de Santiago de Chile, Av. Victor Jara 3659, Santiago, Chile \\
$^{4}$Center for Interdisciplinary Research in Astrophysics and Space Exploration (CIRAS), Universidad de Santiago de Chile, Santiago, Chile \\
$^{5}$European Southern Observatory, Alonso de Cordova 3107, Casilla 19001, Vitacura, Santiago, Chile \\
$^{6}$Joint ALMA Observatory, Avenida Alonso de Córdova 3107, Vitacura 7630355, Santiago, Chile \\
$^{7}$National Radio Astronomy Observatory, 520 Edgemont Road, Charlottesville, VA 22903-2475, USA \\
$^{8}$Massachusetts Institute of Technology, Kavli Institute for Astrophysics and Space Research, Cambridge, MA, 02138, USA \\
$^{9}$Kavli Institute for Astronomy and Astrophysics, Peking University, Yiheyuan Lu 5, Haidian Qu, 100871 Beijing, People's Republic of China \\
$^{10}$Institute for Astronomy, University of Hawaii at Manoa, Woodlawn Drive, Honolulu, HI, 96822, USA \\
$^{11}$Departamento de Ciencias, Facultad de Artes Liberales, Universidad Adolfo Ibáñez, Av. Padre Hurtado 750, Viña del Mar, Chile \\
$^{12}$Escuela de Ciencias, Universidad Viña del Mar, Viña del Mar, 2572007, Chile \\
$^{13}$Núcleo Milenio de Formación Planetaria - NPF, Universidad de Valparaíso, Av. Gran Bretaña 1111, Valparaíso, Chile \\
$^{14}$Univ. Grenoble Alpes, CNRS, IPAG, 38000 Grenoble, France \\
$^{15}$Departamento de Astronom\'{\i}a, Universidad de Chile, Casilla 36-D, Santiago, Chile \\
$^{16}$Facultad de Ingenier\'ia y Ciencias, Universidad Adolfo Ib\'a\~nez, Av. Diagonal las Torres 2640, Pe\~{n}alol\'{e}n, Santiago, Chile \\
$^{17}$Data Observatory Foundation, Chile \\
$^{18}$Department of Physics and Astronomy, University of Nevada, Las Vegas, 4505 S. Maryland Parkway, Las Vegas, NV 89154, USA
\\
$^{19}$Nevada Center for Astrophysics, University of Nevada, Las Vegas, 4505 South Maryland Parkway, Las Vegas, NV 89154, USA  \\
$^{20}$Zentrum für Astronomie der Universität Heidelberg, Institut für Theoretische Astrophysik, Albert-Ueberle-Straße 2,
69120 Heidelberg, Germany
}

\date{Accepted 2023 May 24; Received 2023 May 19; in original form 2023 January 25}

\pubyear{2023}

\begin{document}
\label{firstpage}
\pagerange{\pageref{firstpage}--\pageref{lastpage}}
\maketitle


\begin{abstract}
Episodic accretion is a low-mass pre-main sequence phenomenon characterized by sudden outbursts of enhanced accretion. These objects are classified into two: protostars with elevated levels of accretion that lasts for decades or more, called FUors, and protostars with shorter and repetitive bursts, called EXors. HBC 494 is a FUor object embedded in the Orion Molecular Cloud. Earlier Atacama Large (sub-)Millimeter Array (ALMA) continuum observations showed an asymmetry in the disk at 0\farcs2 resolution. Here, we present follow-up observations at $\sim$0\farcs03, resolving the system into two components: HBC 494 N (primary) and HBC 494 S (secondary). No circumbinary disk was detected. Both disks are resolved with a projected separation of $\sim{0\farcs18}$ (75 au). Their projected dimensions are 84$\pm$1.8 $\times$ 66.9$\pm$1.5 mas for HBC 494 N and 64.6$\pm$2.5 $\times$ 46.0$\pm$1.9 mas for HBC 494 S. The disks are almost aligned and with similar inclinations. The observations show that the primary is $\sim$5 times brighter/more massive and $\sim$2 times bigger than the secondary. We notice that the northern component has a similar mass to the FUors, while the southern has to EXors. The HBC 494 disks show individual sizes that are smaller than single eruptive YSOs. In this work, we also report $^{12}$CO, $^{13}$CO, and C$^{18}$O molecular line observations. At large scale, the $^{12}$CO emission shows bipolar outflows, while the $^{13}$CO and C$^{18}$O maps show a rotating and infalling envelope around the system. At a smaller scale, the $^{12}$CO and $^{13}$CO moment zero maps show cavities within the continuum disks' area, which may indicate continuum over-subtraction or slow-moving jets and chemical destruction along the line-of-sight.
\end{abstract}

\vspace{-5pt}
\section{Introduction}

\label{sec:intro}
During their evolution, young stellar objects (YSOs) dissipate their envelopes while feeding their developing protostars through accretion via a disk. However, YSOs are underluminous compared to the luminosity and accretion rates expected from steady disk accretion. This discrepancy has been established as the ``luminosity problem" \citep{kenyon1990, evans2009}. One potential solution to the luminosity problem is that young stars undergo episodes of high accretion interspersed by quiescent phases. During the episodes of enhanced accretion, large amounts of material are accreted very quickly (e.g., \citealp{dunham2012}). Such phenomena have been observed in low-mass pre-main sequence stars like FU Orionis and EX Lupi, the prototypes of the two subclasses that display the episodic accretion phenomenon. 

FU Orionis type stars, a.k.a. FUors, are pre-main sequence low-mass stars that show variability in both luminosity and spectral type due to variation in the accretion mass rate, on a short timescale \citep{herbig1966}. Their optical brightness can dramatically increase due to enhanced mass accretion by more than five magnitudes over a few months. After this ``outburst'', FUors can remain in this active state for decades. This occurrence is considered episodic and suspected to be common at the early stages of star formation. Throughout an outburst, the star can accrete $\sim$ 0.01 M$_{\odot}$ of material, roughly the mass of a typical T-Tauri disk \citep{andrews2005}. The bolometric luminosity of the FUors during the outburst is 100--300 $L_{\odot}$ and the accretion rate is between $10^{-6}$ and $10^{-4}$ M$_{\odot}$ yr$^{-1}$ \citep{audard2014}. EX Lupi type stars, a.k.a. EXors, are a scaled-down version of the FUors, with shorter and less intense outbursts \citep{herbig2007}. The EXors enhanced accretion stage can last months to years, with accretion rates ranging from 10$^{-7}$ to 10$^{-6}$ M$_{\odot}$ yr$^{-1}$, which are the order up to 5 magnitudes brighter than quiescent periods. The episodic recurrence is also on the order of years (e.g., \citealp{2018A&A...614A...9J, 2022ApJ...929..129G}).

The mechanisms producing the eruptions in systems like FUors and EXors have yet to be understood \citep{cieza2018}. Several different triggers have been proposed to explain this phenomenon, such as disk fragmentation \citep{vorobyov2005, vorobyov2015, zhu2012}, coupling of gravitational and magneto-rotational instability (MRI) \citep{armitage2001, zhu2009}, and tidal interaction between the disk and a companion \citep{bonnell1992,lodato2004, borchert2022}. In terms of evolution, a scenario where FUors are understood as an earlier phase followed by an EXor phase could explain some of the observed properties of these systems \citep{cieza2018}. EXors have less prominent (or lack of) outflows, and smaller masses/luminosities than FUors since the mass loss, accretion, and outbursts will significantly remove the gas and dust material during their evolutionary stages.  

If most stars undergo FUor/EXor-like episodic accretions during their evolution, imaging the circumstellar disks of FUors at sub-mm/mm wavelengths can inform or constrain the underlying outburst mechanisms. For class 0/I objects (e.g., those still accreting from their circumstellar disks and envelopes), the massive disk could be expected to be prone to be gravitationally unstable, which can, consequently, trigger the MRI and/or disk fragmentation. Additionally, close encounters of stars can shape both gas and dust disk morphology \citep{2019MNRAS.483.4114C}. Inferring which scenario plays in each disk requests an analysis that strongly depends on how well-resolved and which features one can extract from the observations. Consequently, when eruptive disks give a hint of irregular morphology or kinematics, follow-up observations at higher resolution are needed to resolve the substructures and infer the nature of the system.

HBC 494 is a Class I protostar, located in the Orion molecular cloud at a distance of 414$\pm$7 pc \citep{2007A&A...474..515M}, and has been classified as a FUor\footnote{The FU Ori classification is still controversial and has been contested by \citet{connelley2018}.}. This young eruptive object (also called Reipurth 50) was discovered during an optical survey and described as a bright, conical, and large nebula reflecting light from a 1.5 arcmin away from a 250 L${\odot}$ infrared source. Both were located at the southern part of the L1641 cloud in Orion \citep{reipurth1985,reipurth1986}. This nebula was claimed to appear after 1955, and its first detailed study showed high variability between 1982 and 1985, a consequence of a primordial IR source variability. Posterior studies have also confirmed its variability. For example, \citet{chiang2015} reported a dramatic brightening (thus clearing of part of the nebula) that occurred between 2006 and 2014. These events can be explained by an episode of outflow coming from HBC 494. More recently, \citet{2019A&A...631A..30P}, using archival photometry data along with Herschel and Spitzer spectra presented the detection of several molecular lines and the spectral energy distribution (SED) of HBC 494. The SED presented strong continuum emission in the mid and far-infrared, which is indicative of envelope emission. Such violent outflows were observed and described in \citet{ruizrodriguez2017}.

ALMA Cycle-2 observations in the millimeter continuum ($\sim$0\farcs25 angular resolution) show that the disk is elongated with an apparent asymmetry \citep{cieza2018}, indicating the presence of an unresolved structure or a secondary disk. In this work, we present ALMA Cycle 4 observations at $\sim$0\farcs03 angular resolution that reveal the binary nature of the HBC 494 system. Only Cycle 4 data were used in this work. The paper is structured as follows. The ALMA observations and data reduction are described in section \ref{sec:obs}. The results from the continuum and line analysis are described in section \ref{sec:results}. The discussion is presented in section \ref{sec:dis}, while we conclude with a summary of our results in section \ref{sec:sum}. 

\section{Observations and data reduction}
\label{sec:obs}

HBC 494 has been observed during the ALMA Cycle-4 in band 6 (program 2016.1.00630.S: PI Zurlo) for two nights, one on the 9\textsuperscript{th} October 2016 and the other on the 26\textsuperscript{th} September 2017 (see table \ref{table}). On the first night, the short baseline configuration was acquired, with a precipitable water vapor of 0.42 mm. The ALMA configuration of antennas was composed of 40 12-m antennas with a baseline from 19 to 3144 m. The flux calibrator and bandpass calibrator were both J0522-3627, and the phase calibrator was J0607-0834. The total integration on the target was 8.39 min. The second night, the long baseline dataset, had precipitable water vapor of 1.08 mm. The ALMA configuration consisted of 40 12-m antennas with a baseline from 42 m to 14851 m. The flux calibrator was J0423-0120, the bandpass calibrator was J0510+1800, and the phase calibrator was J0541-0541. The total integration on the target was 25.55 min. 
\begin{table}
\centering
\caption{List of the Cycle 4 ALMA observations of the target HBC 494, used in this work}
\begin{tabular}{cccccc}

PI & Proj. ID                                     & Ang. Res.                                 & Date of obs. & Integration \\
& & (beam minor axis) & & & \\
\hline
\hline
 Zurlo & 2016.1.00630.S
 & 0\farcs142               
 & 09/10/2016            & 8.39 min                         \\
Zurlo & 2016.1.00630.S
 & 0\farcs027               
 & 26/09/2017            & 25.55 min                         \\
\end{tabular}
\label{table}
\end{table}

The central frequencies of each spectral window were: 230.543 (to cover the transition $^{12}$CO J=2-1), 233.010 (continuum), 220.403 ($^{13}$CO J=2-1), 219.565 (C$^{18}$O J=2-1), and 218.010 (continuum) GHz. The minimum spectral resolution achieved (second night) for $^{12}$CO was 15.259 kHz ($\sim$ 0.02 km/s), while for $^{13}$CO and C$^{18}$CO, it was 30.518 kHz ($\sim$ 0.04 km/s). Both continuum bands presented 2 GHz bandwidths. The data were calibrated with the Common Astronomy Software Applications package \citep[\texttt{CASA} v.5.5,][]{2007ASPC..376..127M}, and the python modules from CASA API, \textit{casatasks} and \textit{casatools} \citep{2022PASP..134k4501C}. The standard calibrations include water vapor calibration, temperature correction, and phase, amplitude, and bandpass calibrations.

\subsection{Continuum imaging}
\label{cont_imag}
We started our analysis by fixing the visibilities' phase center on J2000 05h40m27.448s -07d27m29.65s. The image synthesis of the 1.3 mm continuum emission was performed with the \textit{tclean} task of CASA. For the high-resolution data, we used the Briggs weighting scheme with a robust parameter of 0.5, resulting in a synthesized beam size of 41.2 $\times$ 29.8 mas, and a position angle of 40.6. The pixel scale of the image was set to 1.5 mas. One iteration step of self-calibration was applied to both observations. For the first night (low-resolution image: 142 mas), it resulted in an improvement of $\sim$6 SNR. For the second night (high-resolution image: 27 mas), it resulted in an SNR improvement of $\sim$ 1.6. 

Our final continuum image was produced from a concatenated measurement set combining both, the short and the long baseline visibilities, using the \textit{concat} task of CASA. We used the same image synthesis parameters used for the second night of observation.

\subsection{Line imaging}

\begin{figure}
\begin{center}
    \includegraphics[width=0.49\textwidth]{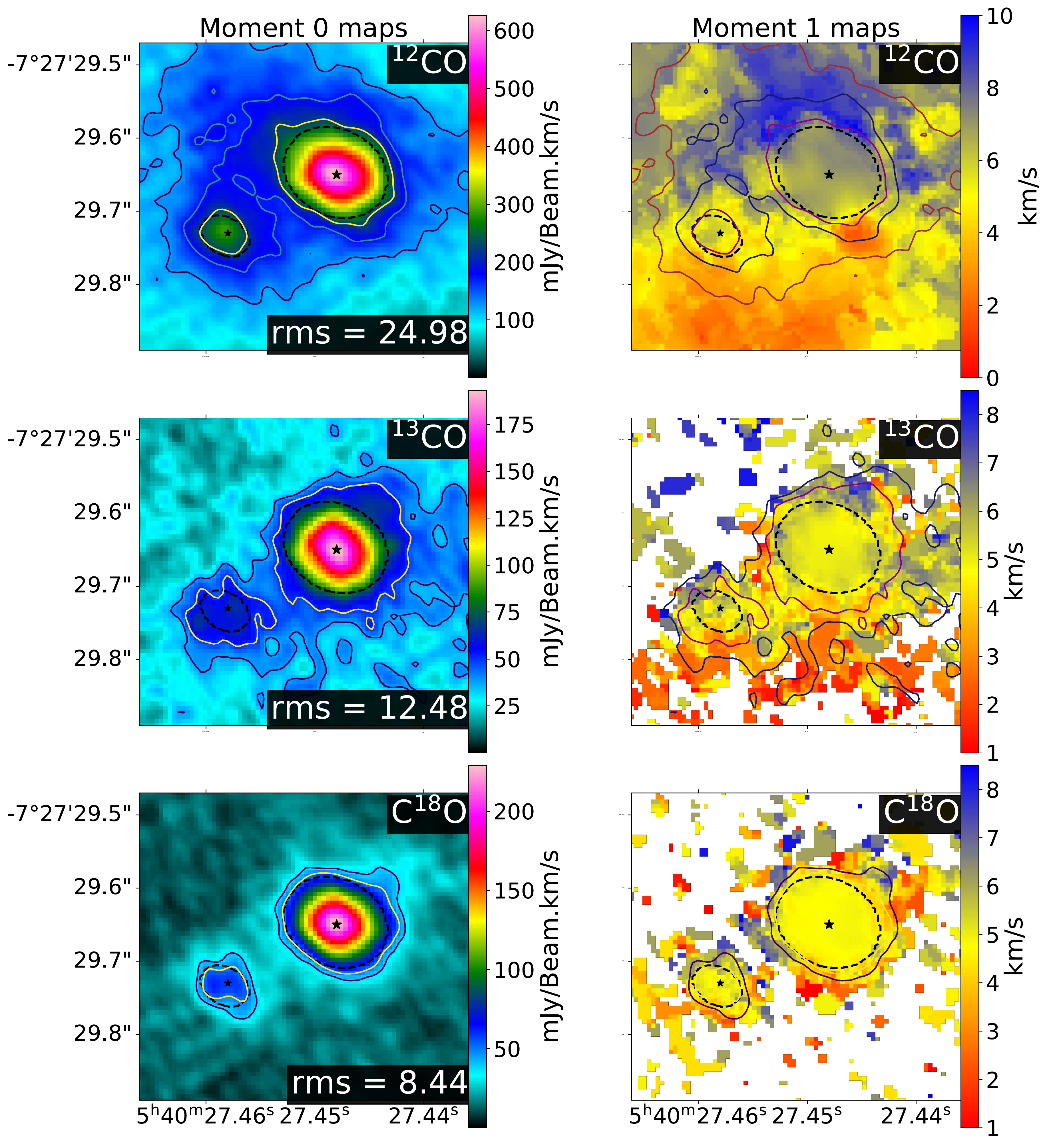}
    \caption{Small-scale moment maps for the analyzed CO isotopologues, $^{12}$CO (top), $^{13}$CO (middle) and C$^{18}$O (bottom), without subtraction of the continuum. The black stars mark the peak flux positions of the continuum disks. The dashed black contours correspond to 100$\times$rms from the continuum emission (100$\times$34$\mu$Jy/beam). Left: Moment 0 maps for each mentioned molecular line. The rms values at the bottom of each panel are in units (mJy/beam)(km/s). The $^{12}$CO moment 0 map (top-left) has contour levels of 5, 7, and 9$\times$rms showed at the bottom of the panel. $^{13}$CO and C$^{18}$O moment 0 maps (middle and bottom - left) have contour levels of 3 and 4$\times$rms showed at the bottom of each panel. Right: Moment 1 maps for each mentioned molecular lines. The contours are the same as each left panel respectively. The maps include only pixels above 3 times the rms measured from the set of channels used for each molecular line (-6 km/s to 18 km/s for $^{12}$CO and 0 km/s to 11 km/s for $^{13}$CO and C$^{18}$O).}
    \label{moments_small_with_cont}
\end{center}
\end{figure}

The visibilities were fixed at the same phase center as was done for the continuum coordinates (see subsection \ref{cont_imag}). After producing a dirty image from the visibilities, we noticed that the emission mimicked the presence of gaseous envelopes around the disks (moment 0 maps, see Figure \ref{moments_small_with_cont}, left column) but no gas dynamic signatures were revealed (moment 1 map, see Figure \ref{moments_small_with_cont}, right column). Therefore, we proceeded to remove the visibility's continuum contributions by using the CASA task \textit{uvcontsub}. 

Next, we performed the \textit{tclean} process using Briggs weighting with a robust parameter of 0.5, Hogbom deconvolver, and pixel scale of 6 mas. The total velocity ranged between -6 km/s to 18 km/s, although a smaller range (0 km/s to 11 km/s) was used to create the $^{13}$CO and C$^{18}$O moment maps. We used spectral resolution widths of 1 km/s for $^{12}$CO and of 0.3 km/s for $^{13}$CO and C$^{18}$O. The rms values, obtained for the channel without a clear signal on the channel maps (4 km/s), were 1.9, 3.6, and 2.6 mJy/beam for $^{12}$CO, $^{13}$CO and C$^{18}$O, respectively.

\section{Results}
\label{sec:results}
\subsection{Continuum analysis}
\label{sec:con}

\begin{figure*}
\includegraphics[width=1\textwidth]{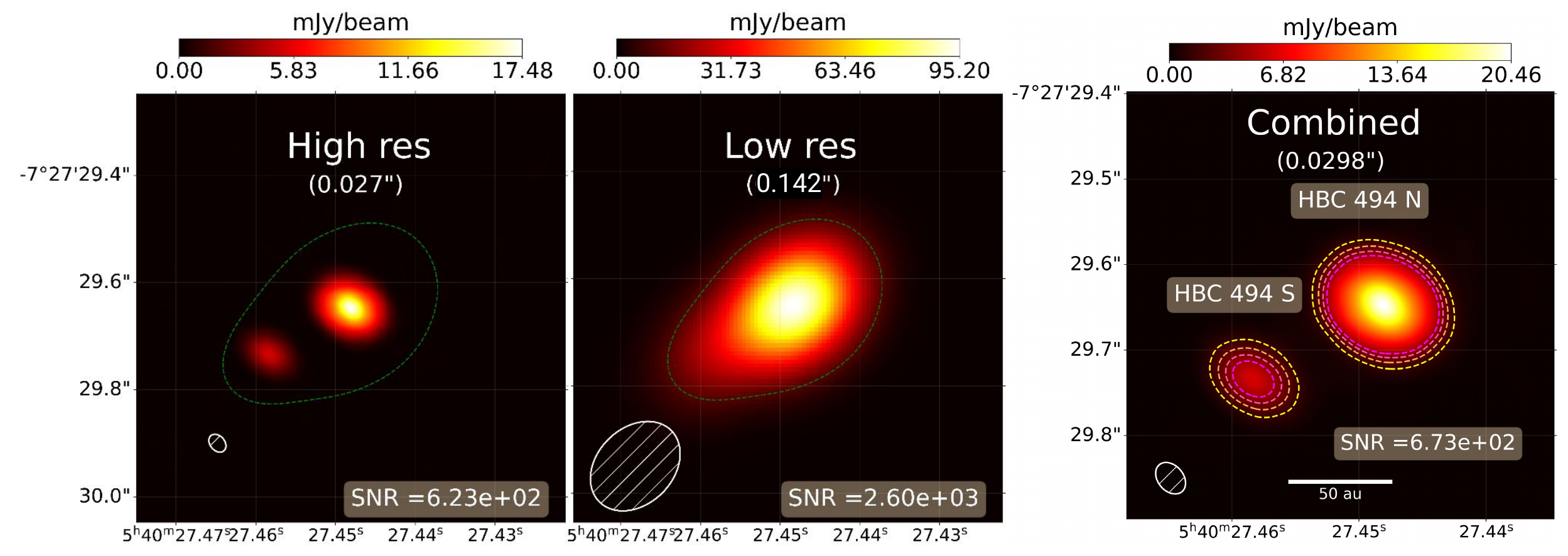}
\caption{ALMA observations at 1.3 mm of the HBC 494 system. The images reveal the binary nature of HBC 494 with two components surrounded by continuum emission associated with circumstellar dust. The respective signal-to-noise ratios of each image are shown in each panel. We show the resulting images from the long baseline observation with a resolution of 0\farcs027 arcsecond (\textit{left}), the short baseline data at 0\farcs142 arcsecond resolution (\textit{middle}), and the combination of the two observations with 0\farcs0298 (\textit{right}). Dashed green lines in the left and middle images represent a 3-sigma contour associated with the lower-resolution continuum image. The combined image has a beam size of $\sim$ 41 x 30 mas, shown in the lower-left corner of the image. Its contours correspond, from the yellow (more external) to the pink dashed lines, to 10, 40, 70, and 100 times the rms value, respectively. The rms is $\sim${0.034 mJy/beam}.}
\label{stages_cont}
\end{figure*}

With the high-resolution data shown in Figure~\ref{stages_cont}, we could reveal the binary nature of the HBC 494 system. Each of the components in HBC 494 is surrounded by continuum emission, most likely associated with circumstellar disks.

\begin{figure*}
\begin{center}
\includegraphics[width=1\textwidth]{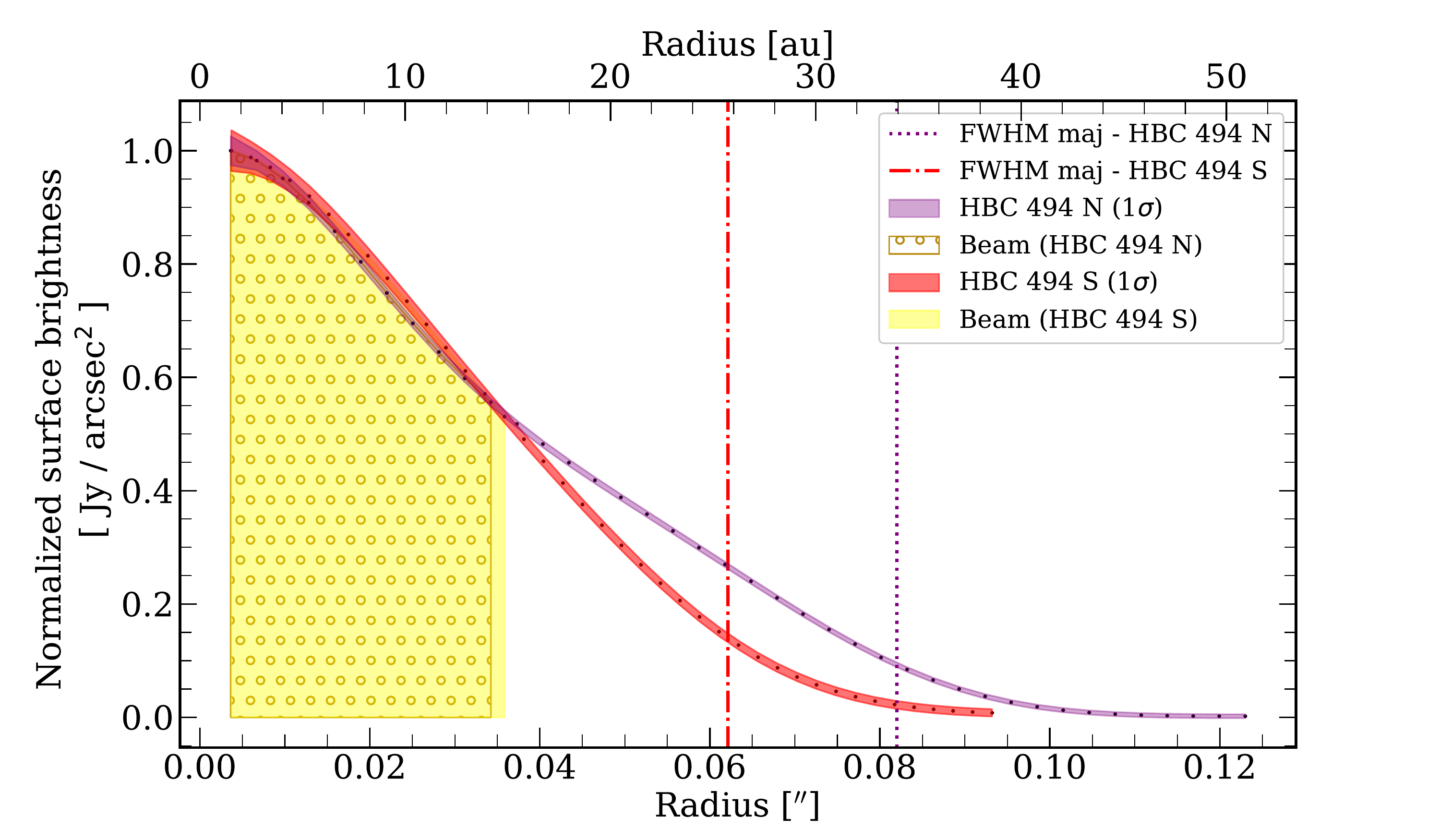}

\caption{Radial brightness profile of HBC 494 N and HBC 494 S. The surface brightness (dots), measured in apertures ranging from the beam's major axis to 3 times the semi-major axis FWHM of each continuum disk, were normalized to the peak fluxes. The shaded regions indicate 1$\sigma$ error measured in each aperture. The dashed lines correspond to the ALMA imfit FWHM of each disk.}

\label{rad_prof}
\end{center}
\end{figure*}

We performed a 2D Gaussian fitting with the \textit{imfit} tool of CASA in order to characterize the newly resolved components. The projected separation between the two sources is 0\farcs18 (75 au). We name each component with the usual convention, i.e. ``N'' and ``S'' components, referring to the northern and the southern source, respectively. We assume the northern component to be the primary as it is five times brighter than the southern (secondary) disk. Both individual disks are resolved according to our Gaussian fit. For the primary component, we find a major axis FWHM of 84.00 $\pm$ 1.82 mas and a minor axis of 66.94 $\pm$ 1.50 mas (34.8 $\pm$ 0.7 $\times$ 27.8 $\pm$ 0.6 au), with a position angle of 70.01 $\pm$ 4.46 deg (values deconvolved from the beam). For the secondary, the major axis FWHM is 64.60 $\pm$ 2.49 mas and the minor axis FWHM is 45.96 $\pm$ 1.89 mas (26.7 $\pm$ 1.0 $\times$ 19.9 $\pm$ 0.8 au), and the position angle is 65.38 $\pm$ 5.32 deg. The integrated fluxes for ``N'' and ``S'' are 105.17 $\pm$ 1.89 mJy and 21.06 $\pm$ 0.63 mJy, respectively. The peak fluxes are 22.96 $\pm$ 0.35 mJy/beam and 8.71 $\pm$ 0.21 mJy/beam for the primary and secondary sources, respectively. The rms is $\sim$0.034 mJy/beam, as observed in a circular region with a radius of 225 mas without emission. The inclinations are calculated using the aspect ratios of the disks. These are assumed as projected circular disks when face-on, thus elliptical when inclined:  
\begin{center}
    $i = \arccos{\left( \frac{b}{a} \right)}$,
\end{center}
\noindent where ``b" and ``a" are the semi-minor and semi-major axes, respectively. It resulted in 37.16 $\pm$ 2.36 degrees inclination for HBC 494 N and 44.65 $\pm$ 3.27 for HBC 494 S. At 30 mas resolution, no substructures were detected (see the radial profiles in Figure~\ref{rad_prof}).  

A large fraction of the mm-sized grains in protoplanetary disks resides in the midplane. They are usually optically thin to their own radiation. Thus, assuming standard disks, we can use the observed fluxes to estimate the dust masses of HBC 494 N and S. For that, we use the following formula \citep[as in][]{beckwith1990, andrews2005, cieza2018}: 

\begin{equation}
    M_{\rm dust} =\frac{F_{\nu}d^{2}}{k_{\nu}B_{\nu} (T_{\rm dust})},
\end{equation}
\noindent where $F_{\nu}$ is the observed flux, \textit{d} is the distance to the source, $B_{\nu}$ is the Planck function and $\kappa_{\nu}$ is the dust opacity. Adopting the distance of 414 pc, isothermal dust temperature ($T_{\rm dust}$ = 20 K) and dust opacity assuming a $M_{\rm gas}/M_{\rm dust}$ fraction of 100, ($\kappa_{\nu}$ =  10 ($\nu /$ 10$^{12}$Hz) cm$^{2}$ g$^{-1}$; \citealt{beckwith1990}) we get dust masses of 1.43 $M_{\rm Jup}$ (HBC 494 N) and 0.29 $M_{\rm Jup}$ (HBC 494 S). Following the standard gas-to-dust mass ratio of 100, we report disk gas masses estimations of 143.46 $M_{\rm Jup}$ for HBC 494 N, and 28.68 $M_{\rm Jup}$ for HBC 494 S. The physical parameters based on the CASA analysis are listed in Table \ref{compiled_disks}.

\begin{table}
\centering
\caption{ALMA imfit analysis results and physical parameters inferred from the HBC 494 continuum disks.}
\label{compiled_disks}

\small\addtolength{\tabcolsep}{-2pt}
\begin{tabular}{lcc}
                                        & HBC 494 N             & HBC 494 S            \\
\hline
\hline
Flux density (mJy)                      & 105.2 $\pm$ 1.9  & 21.1 $\pm$ 0.6 \\
Peak flux (mJy / beam)  & 23.0 $\pm$ 0.4 & 8.7 $\pm$ 0.2                        \\
Major axis (mas)                        & 84.0 $\pm$ 1.8  & 64.6 $\pm$ 2.5 \\
Minor axis (mas)                        & 66.9 $\pm$ 1.5  & 46.0 $\pm$ 1.9 \\
Major axis (au)                        & 34.8 $\pm$ 0.7  & 26.7 $\pm$ 1.0 \\
Minor axis (au)                        & 27.8 $\pm$ 0.6  & 19.9 $\pm$ 0.8 \\
Position angle (deg)                    & 70.0 $\pm$  4.5 & 65.4 $\pm$ 5.3 \\

Beam major axis (mas)    & 41.2 & 41.2 \\
Beam minor axis (mas)    & 29.8 & 29.8  \\
Beam Position angle (deg)               & 40.6                         & 40.6                        \\
Beam area (sr)                          & 3.3e-14                      & 3.3e-14   \\
rms ($\mu$Jy / beam)                          & 34.0                      & 34.0 \\
\hline
Dust mass $\, (M_{\rm Jup})$ & 1.4 & 0.3 \\
Gas mass $\, (M_{\rm Jup})$ & 143.5 & 28.7 \\
Inclination (deg) & 37.2$\pm$2.4 & 44.7$\pm$3.3 \\
FWHM Radius (au) & 34.8$\pm$0.7 & 26.7$\pm$1.0 \\
\end{tabular}%
\end{table}

\subsection{Line analysis}
\label{sec:gas}

The next two subsections describe the large and small-scale structure analyses. The large scale, using the extent of 8000 au, shows the results of outflows and envelopes. At such a scale, there is not enough resolution to display both disks' dynamics and their possible interactions. The small scale (150 au) fulfills this role, and thus, both gas scenarios must be taken into account.

\subsubsection{Large-scale structures (8000 au)}

\begin{figure*}
\begin{center}
\includegraphics[width=1\textwidth]{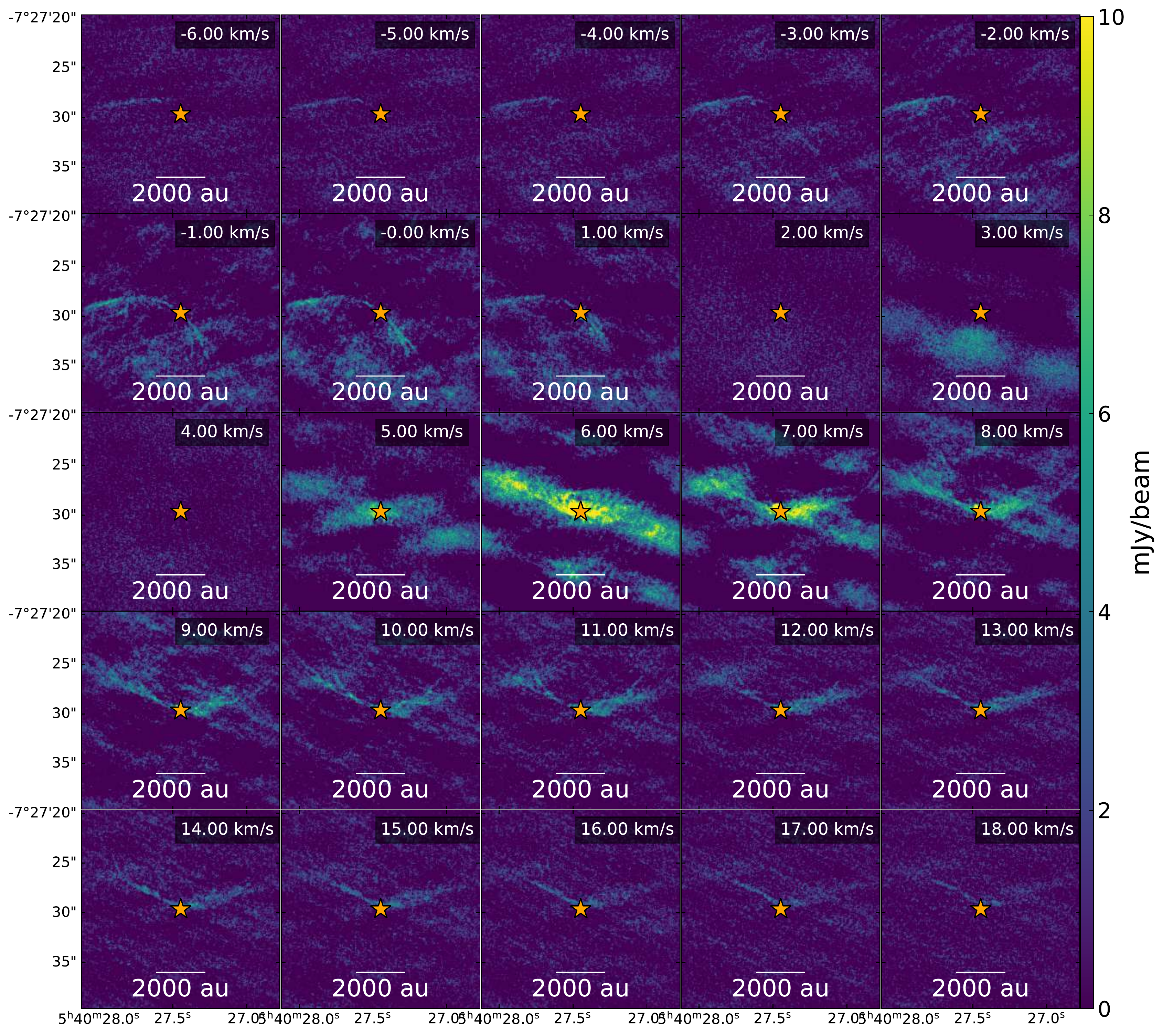}
\caption{$^{12}$CO channel maps of the HBC 494 system. The star in the center marks the position of the continuum disks. The southern arc is more evident between the velocities -6 km/s to 1 km/s, while the northern is more easily seen between 7 km/s and 18 km/s.}
\label{12cochannel}
\end{center}
\end{figure*}

\begin{figure*}
\begin{center}
\includegraphics[width=1\textwidth]{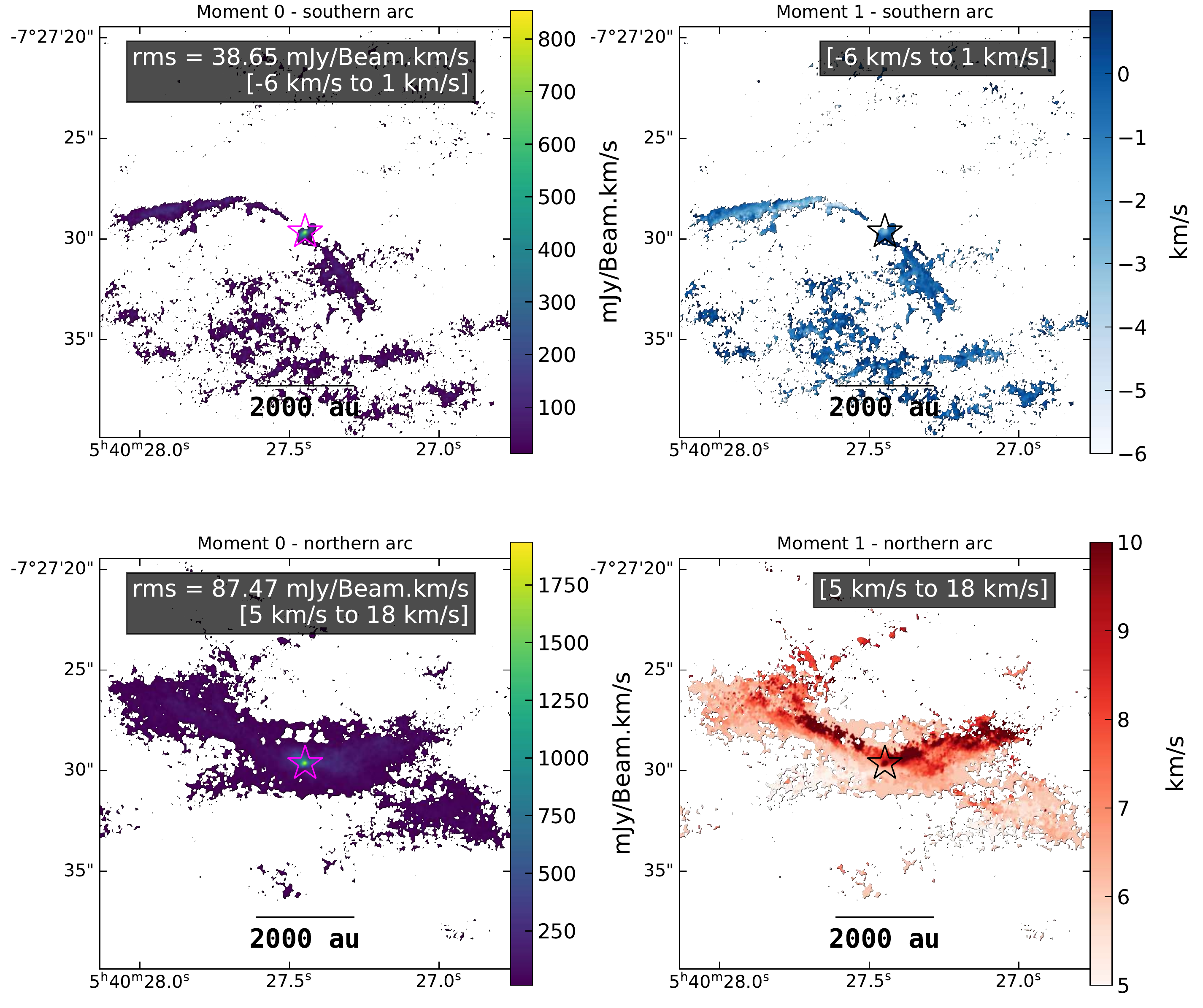}
\caption{Large-scale $^{12}$CO moment maps split in each one of the bipolar outflow structures. The left column corresponds to moment 0 maps, and the right column to moment 1 maps. The plots in the same row refer to the same structure, which is the southern arc (\textit{upper row}), and the northern arc (\textit{bottom row}).  The position of the HBC 494 N is marked with a star. The maps include only pixels above 2 times the rms measured from the set of channels shown at the top of each panel.}
\label{arcs}
\end{center}
\end{figure*}

\begin{figure*}
\includegraphics[width=1\textwidth]{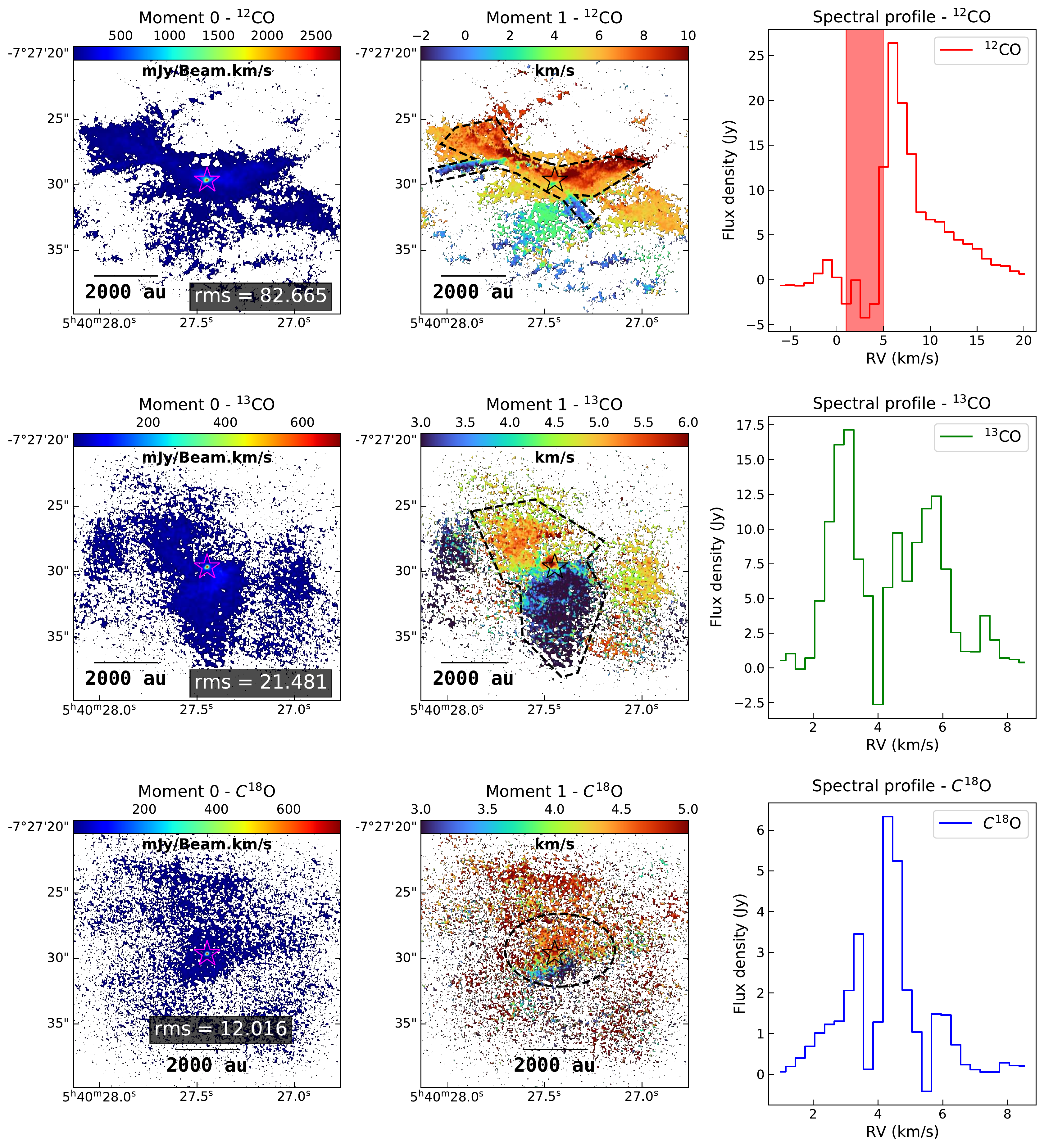}
\caption{Large-scale moment maps for the analyzed CO isotopologues and spectral fluxes. The left column corresponds to moment 0 maps, the middle one to moment 1 maps, and the third column to spectral profiles. The rms values at the bottom of each panel in the left column are in units (mJy/beam)(km/s). The plots in the same row refer to the same molecular lines, which are $^{12}$CO (\textit{upper row}), $^{13}$CO (\textit{middle row}), and C$^{18}$O (\textit{bottom row}). The dashed black lines in the moment 1 column correspond to the regions used to produce the spectral profiles. The red shaded area in $^{12}$CO profile represents the channels we excluded before creating the moment maps lying in the same row (to avoid cloud contamination). The center of stars, in the middle of each figure, marks the position of both disks. The maps include only pixels above 2 times the rms measured from the set of channels used for each molecular line (-6 km/s to
18 km/s for 12CO and 0 km/s to 11 km/s for 13CO and C18O).}
\label{largescalemoments}
\end{figure*}

The channel maps (Figures \ref{12cochannel}, \ref{13cochannel} and \ref{c18ochannel}) show southern and northern $^{12}$CO wide-angle arcs ($\sim150^{\circ}$) and no clear signal of large structures from the other molecular lines. This bipolar outflow was previously described in \citet{ruizrodriguez2017}. The wide angular structure shows an outflow morphology expected for Class I disks \citep{arce2006}. In our observations, the southern and northern arcs are defined by the velocity ranges of [-6 km/s to 1 km/s] and [5 km/s to 18 km/s], respectively. The moment 0 and 1 $^{12}$CO arcs are shown in Figure~\ref{arcs}. However, it is noticeable that the northern arc (redshifted emission) dominates the displayed velocities. The lack of signal in the southern arc can be explained by lower molecular density, or due to intracloud absorption. Figure~\ref{largescalemoments} shows the three molecular lines' concentration and dynamics (moment 0 and 1 maps), as well as the spectral profile measured within the dashed regions. 

The $^{13}$CO emission traces the rotating, infalling, and expanding envelope surrounding the system, showing a small deviation from the $^{12}$CO outflowing arcs' rotation axis. The blue-shifted emission concentrates around 3 km/s, while the red-shifted emission is around 5 km/s. It is noticeable that the $^{13}$CO emission has a rotation axis almost perpendicular to the $^{12}$CO outflows axis, ensuring that the physical processes behind the two are different. The spectral $^{13}$CO also shows a dip around 4 km/s, but this may not trace an absorption but a lack of signal due to the analyzed uv-coverage or cloud contamination.
 
The C$^{18}$O emission, however, is faint and hardly distinguishable from the surrounding gas in moment 0 maps. It has also the lowest abundance of the three analyzed isotopologues, being a good tracer of the higher gas densities in the innermost regions of the cloud. Its velocity maps show that a faint blueshifted motion was detected around 3 km/s, very weak when compared to the redshifted emission detected principally around 5 km/s. Due to the higher concentration of gas in the northern part, it is reasonable that we observed a weak gas counterpart in the southern region. At large-scale, our results are comparable with the ones presented in \citep{ruizrodriguez2017}. Furthermore, a lower limit of the envelope material of $\sim$600 Jy.km/s towards HBC 494 could be traced using Total Power (TP), ALMA compact array (ACA), and the main array ALMA observations with C$^{18}$O data in the range of 4.3$\pm$0.5 km s$^{-1}$ (Ru\'iz-Rodr\'iguez private communication).

\subsubsection{Small scale structures (150 au resolution)}

We started by removing the continuum contribution from all molecular lines observed ($^{12}$CO, $^{13}$CO, C$^{18}$O). We already expected that $^{12}$CO would trace the gas in larger scales due to its optical depth and, thus, could blend with smaller gas signatures. To avoid cloud contamination during the creation of $^{12}$CO small-scale moment maps, we removed the channels that were more affected. We considered velocities in the range of -6 km/s to 1 km/s, and from 5 km/s to 18 km/s, as similarly done for the different large-scale arcs described in the previous subsection. 

\begin{figure*}

\begin{center}
    \centering
    \includegraphics[width=1\textwidth]{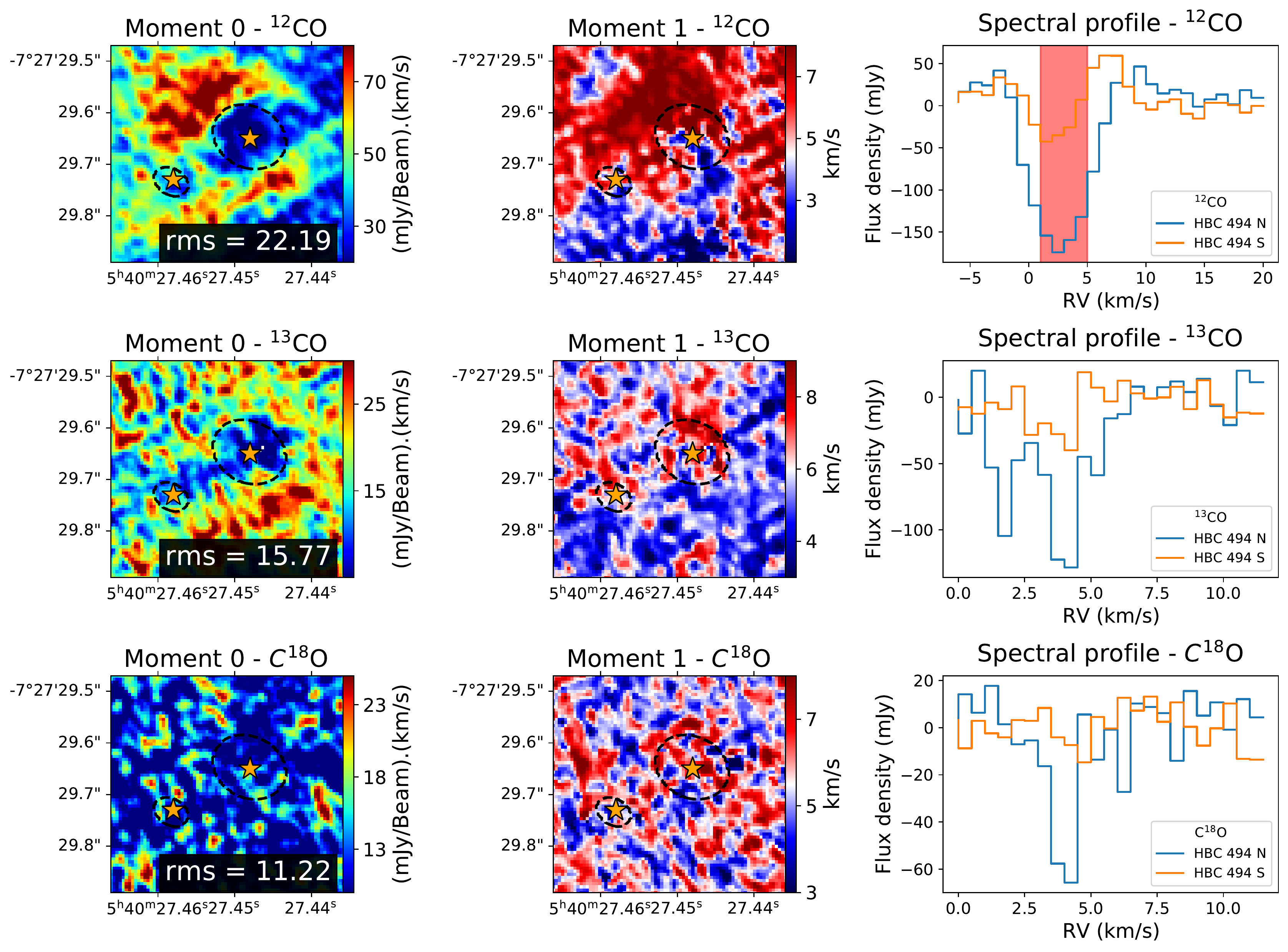}
    \caption{Small-scale moment maps for the analyzed CO isotopologues, after subtraction of the continuum, and their spectral profiles. The first column corresponds to moment 0 maps, and the middle one to moment 1 maps. The rms values at the bottom of each panel in the left column are in units (mJy/beam)(km/s). The molecular lines correspond to $^{12}$CO (\textit{upper row}), $^{13}$CO (\textit{middle row}) and C$^{18}$O (\textit{bottom row}). The dashed black lines spatially correspond to the contour level equal to 100 times the continuum rms (100$\times$34$\mu$Jy/beam), in the continuum combined image (Figure \ref{stages_cont}, \textit{right}). The orange stars mark the peak flux positions of the continuum disks. The right column represents the spectral profiles measured from velocity maps taken within each disk. The red shaded area in $^{12}$CO profile represents the channels we excluded before creating the moment maps lying in the same row (to avoid cloud contamination).}
    \label{moments_small}
\end{center}
\end{figure*}

Following, CO channel maps and moment 0 and 1 maps were produced to explore potential hints of interaction between the stars and the two circumstellar disks. In particular, we looked for disk substructures and perturbed rotation patterns. The moment maps can indeed provide valuable information about the binary orbit and the disks' geometry. However, by looking at the molecular lines in this scale (see the moment maps in Figure~\ref{moments_small}, left and middle columns), we could not clearly detect the presence of the disks or their rotational signatures. We can only notice that the $^{12}$CO and $^{13}$CO moment 0 maps show an interesting pattern, where less gas emission is found where the continuum disks are. Additionally, the area within the continuum disks highlights regions similar to cavities or ``holes" (Figure~\ref{moments_small}, left column, upper and middle row). Interestingly, \citet{ruizrodriguez2022} observed similar features from the same molecular lines, in addition to HCO$^{+}$, for another FUor system, V883 Ori. The work also showed that the origin of cavities and ring emission around disks can be interpreted in two ways, one based on gas removal and the other on optical depth effects. 

If the signal traces the lack of gas surrounding the disks, the leading mechanism may be slow-moving outflows. The central regions of young continuum disks are expected to be constantly carved by the influence of magneto-hydrodynamical jets \citep[][]{frank2014}. In addition, it is also expected that the lower emission observed comes from the chemical destruction by high-energy radiation (more active disks will create bigger cavities). By looking at the size of the gaps, we can observe that HBC 494 S exerts a smaller influence on the gas than HBC 494 N and, thus, HBC 494 N is the more active disk in the system if this hypothesis is correct.

Another interpretation of the negatives in our gas maps is that this is a spurious result of continuum subtraction. Usually, the continuum emission is inferred from channels devoid of gas line emission. The usual procedure does not consider the line emission where these channels overlap. Suppose the emission lines are optically thick above the continuum. In that case, the foreground gas can significantly absorb the photons coming from the underlying continuum, which may lead to a continuum overestimation. Consequently, in such cases, the continuum may be over-subtracted, causing the ``hole" feature that is stronger where the line peaks (see e.g., \citealp{boehler2017,weaver2018}).

\section{Discussion}
\label{sec:dis}
\subsection{Multiplicity and triggering mechanisms}

Besides HBC 494, there are $\sim$ 25 known FUor objects within 1 kpc \citep{audard2014}. As previously commented, many triggering mechanisms can cause episodic accretion events. From this sample, only HBC 494 and a few other FUor systems are known binaries (e.g., FU Orionis, L1551 IRS, RNO 1B/C, AR 6A/B; \citealp{pueyo2012} and references therein). 

Given the high occurrence rate of stellar binaries harboring disks, the lack of detection of multiple YSO systems remains a case of investigation. Additionally, young disks evolve in crowded star-forming regions, enhancing the hypothesis that the outbursts of Class 0/I  disks may affect other disks, driving episodic accretion events. Also, surveys and studies using ALMA and NIR data for Class I-III disks, show that there is a lack of detection when only visual multiple systems (separations of 20-4800 au) are considered (e.g., \citealp{zurlo2020b,zurlo2021}). When all the possible separations are taken into account, the multiplicity frequencies considerably increase. For Taurus, for example, it goes up to 70\% when spectroscopic binaries are included \citep{kraus2011}. In the case of Orion, 30\% of the systems are multiple with separations between 20 and, 10000 au \citep[][]{tobin2022}. The latter study also noticed that the separations decrease with time (when comparing Class 0, I, and flat-spectrum disks) and that the multiplicity frequency in Class 0 is higher than in the later evolutionary stages. Therefore, multiple eruptive young systems must be common, and more detectable within evolutionary time. They probably are not frequently detected due to the short time duration of enhanced accretion events in comparison with the lifetime of disks in class 0/I stages (see \citealt{audard2014} for a FUors review). Still, it is not clear if disks belonging to close-separation systems are more susceptible to eruptive events than isolated ones. 

For each HBC 494 disk, no clue of trigger mechanisms was found, since the gas and continuum data did not reveal clear spiral or clumpy features indicating a case of infall or GI (see e.g., \citealp{zhu2012,kratter2016}). Looking at the dust continuum (Figure \ref{stages_cont}), CO isotopologues moment maps (Figure \ref{moments_small}), and the scattered $^{12}$CO emission (Figure 12 of \citealt{ruizrodriguez2017}), the hypothesis of stellar flybys is also not encouraged since no trace of perturbation was detected (\citealt{cuello2020, Cuello+2023}). Moreover, dynamical data (small-scale moment 1 maps) also can provide clues for the triggering mechanism \citep{vorobyov2021}. However, it requires clear detection of quasi-keplerian rotation, which was not observed. 

In the case of binaries, the first component to ignite the FUor outbursts can quickly trigger the secondary one by inducing perturbations and mutual gravitational interactions \citep{bonnell1992, Reipurth+2004, vorobyov2021, Borchert+2022a, Borchert+2022b}. Therefore, we can assume that all the disks in FUor multiple systems like HBC 494 might have undergone successive enhanced eruptive stages, despite the differences in mass and radius between each component. However, it is not clear how the outbursts in HBC 494 N and HBC 494 S affect each other since the moment maps do not show the connection between disks. Although, on larger scales, we detected the difference in the amount of gas mass detected in the northern regions compared to the southern. This scenario may be described if the more massive disk (i.e. HBC 494 N) has a higher contribution to the outbursts and winds.

With astrometric data, the eccentricity of orbits can be determined. If both disks are still accreting gas from the environment, it is expected that quasi-circular orbits trigger enhanced accretion every few binary periods. Eccentric orbits, on the other hand, can induce eruptive events in every orbit, preferentially when the binaries reach the pericentre (see e.g. \citealp{2015MNRAS.448.3545D,2022arXiv221100028L}). Observational evidence of circumbinary disks in different YSO stages was found (see e.g. \citealp{1994A&A...286..149D,1997AJ....113.1841M,2016Natur.538..483T,2020ApJ...897...59M}) and are seen as a common counterpart of young binaries in formation. However, no HBC 494 circumbinary gas disk was observed due to cloud contamination and optical depth effects.
\subsection{Disk sizes and masses}
FU Orionis, the precursor of the classification FUor, is also a binary system. HBC 494 system, however, has smaller disks separation compared to the 210 au between FU Orionis north and south components \citep[][]{perez2020}. FU Orionis components have similar sizes and masses between their disks, contrary to the HBC 494 components. We can argue that, due to HBC 494 being a close-packed system with substantially different masses between their components, a scenario of radius truncation can be tested. However, the FU Orionis disks are exceptional compared to other FUors, which are generally more massive and larger. Nevertheless, it is important to check if HBC 494 follows the trend which shows that FUor objects are more massive than class 0/I disks and that disks from multiple systems are smaller compared to those from non-multiple systems \citep{cieza2018,hales2020,tobin2020,zurlo2020}. First, we will describe our results regarding masses and radius. Then, we will compare HBC 494 disks to FUors and Orion YSOs in the literature.

\begin{figure*}
\begin{center}
\includegraphics[width=1\textwidth]{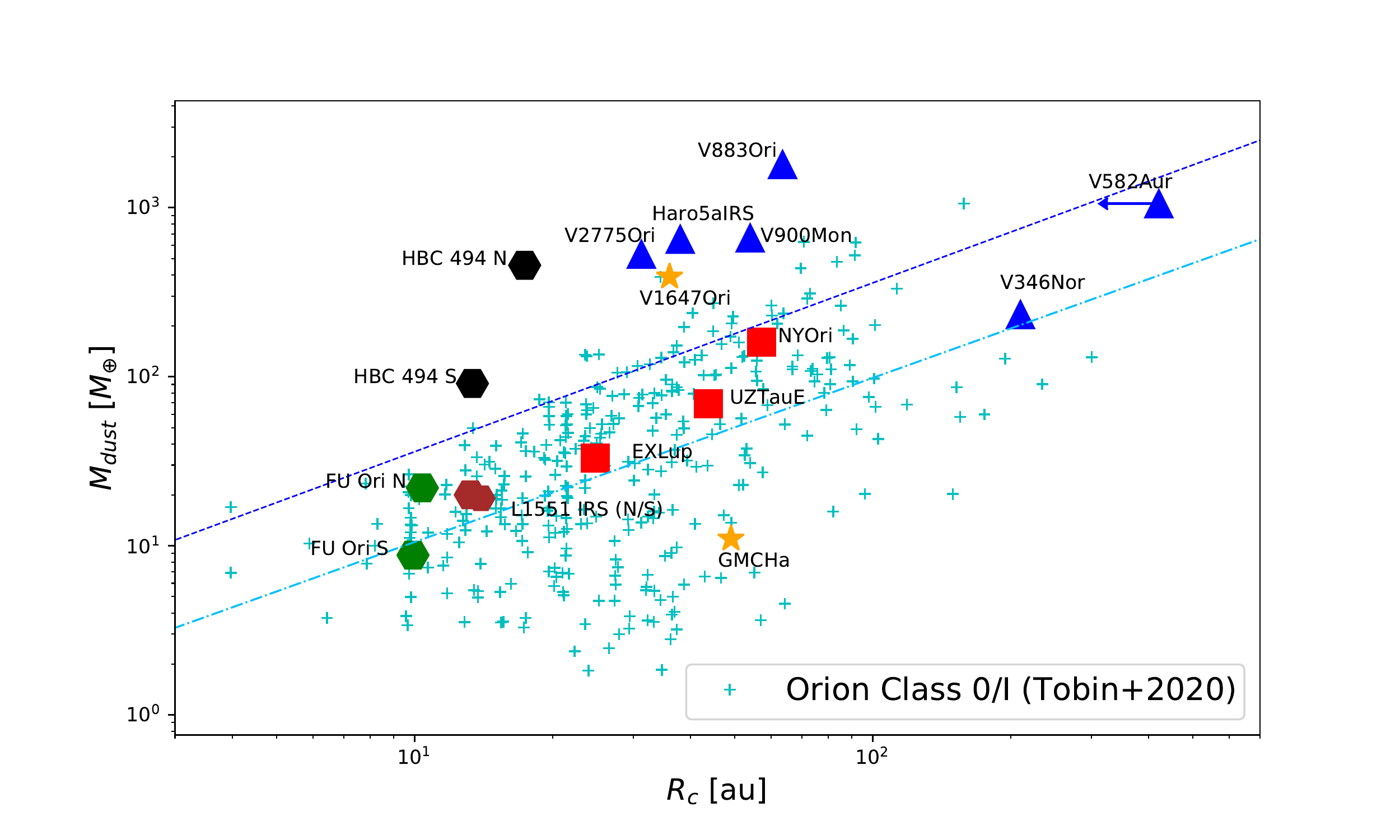}
\caption{Dust disk masses as a function of characteristic radii (FWHM/2) for FUors (blue triangles), EXors (red squares), for double FUor and EXor classification (yellow stars), and Class 0/I objects (cyan crosses), as similarly presented in \citet{hales2020}. The horizontal arrow on V852 Aur denotes the disk radius in an upper limit. The hexagons represent eruptive/multiple systems, with matching colors for disks lying in the same system. The data for HBC 494 disks was generated in this work. The blue dashed line corresponds to the power-fit law to the FU/EXor data (Spearman correlation coefficient: 0.63), and the cyan dash-dotted line, to the Class 0/I data (Spearman correlation coefficient: 0.55). To generate this plot, we obtained data from \citet{kospal2017,hales2018,cieza2018,takami2019,cruz2019,perez2020,hales2020}.}.
\label{radiusxmass}
\end{center}
\end{figure*}

We assumed dust masses (assuming the gas-to-dust ratio of 100) inferred by optically thin approximation (described in section \ref{sec:con}). However, it is worth mentioning that the optically thin analysis can not take into account the innermost regions of the disk (optically thick). This may lead to an underestimation of the dust masses. Also, an underlying miscalculated dust grain temperature can alter the results, overestimating the masses if the dust is warmer than expected. The chosen assumptions of temperatures (fixed 20 K for dust grains) and opacities for both disks lead to 1.43 $M_{\rm Jup}$ and 0.29 $M_{\rm Jup}$ for HBC 494 N and HBC 494 S, respectively. Based on these results, we can say that the HBC 494 N is comparable to other FUor sources, but HBC 494 S has its dust mass comparable to EXor disks \citep{cieza2018}. The disk sizes were calculated using the deconvolved Gaussian FWHM/2 radius obtained from 2D Gaussian fits. This methodology was also used for other ALMA datasets (e.g., \citealp{hales2020, tobin2020}). Therefore, it allows a consistent comparison with other disks. The dust mass vs radius comparison between HBC 494 N and HBC 494 S, a sample of 14 resolved FUors and EXors (extracted from \citealp{hales2020}) and Class 0/I systems \citep[][]{tobin2020}, can be seen in figure \ref{radiusxmass}. Here, as previously stated, we notice that HBC 494 disks (black, hexagon symbols), as other eruptive systems (hexagon, square, and triangle symbols), present higher masses than young systems, which are not in the stage of episodic accretion. However, there is no obvious distinction between the disk sizes of individual systems, eruptive or not. A clear difference is seen when we compare the disks from multiple systems (hexagon symbols) and single systems. The latter present bigger sizes as they are not targets for tidal truncation, enhanced radial drift, and more aggressive photoevaporation --- processes known to rule close-systems dynamics and evolution (see e.g., \citealp{kraus2012,harris2012,rosotti2018,zurlo2020b, zurlo2021}).

\subsection{Binary formation and alignment}

Different scenarios have been proposed to explain 
the formation of binary systems. The main one for close-separated systems is fragmentation, followed by dynamic interactions (for a review, see \citealp{offner2022}). 

The fragmentation process consists of partial gravitational collapse from self-gravitating objects. To be successful, many initial physical conditions are relevant as thermal pressure, density, turbulence, and magnetic fields. In addition, the fragmentation can be divided into two main classes: direct/turbulent (e.g., \citealp{1979ApJ...234..289B,1997MNRAS.288.1060B}) and rotational (e.g., \citealp{larson1972,1994MNRAS.269..837B,1994MNRAS.269L..45B,1994MNRAS.271..999B,1997MNRAS.289..497B}). 

The direct/turbulent fragmentation from a collapsing core is highly dependent on the initial density distribution. It can form wide-separated multiple YSO systems with uncorrelated angular momentum. Thus, the direct/turbulent fragmentation may lead to preferentially misaligned multiple systems \citep{2000MNRAS.314...33B, 2016ApJ...827L..11O, 2018MNRAS.475.5618B, 2019ApJ...887..232L}. The rotational fragmentation, instead, is caused by instabilities in rotating disks and leads to preferentially spin-aligned and coplanar systems \citep{2016ApJ...827L..11O, 2018MNRAS.475.5618B}. Moreover, later dynamical processes such as stellar flybys \citep{ClarkePringle1993, Nealon+2020, Cuello+2023} and misaligned accretion from the environment \citep{Dullemond+2019, Kuffmeier+2020, Kuffmeier+2021} can also induce alignment or misalignment.

The HBC 494 disks have similar inclinations ($\Delta i$ = 7.5 $\pm$ 4.1 degrees) and similar PAs ($\Delta PA$ = 4.6 $\pm$ 7.0 degrees). The relative orientations of the disks suggest that the system is coplanar. Assuming they are quasi-coplanar, with the same PA, the unprojected separation would be $\sim$0\farcs24 (99 au). Therefore, the HBC 494 disks may show a spin-alignment situation, more typical to observed close-separated systems (see, e.g., \citealp{2016ApJ...818...73T,tobin2020}). Thus, we tentatively suggest that HBC 494 was formed by rotational disk fragmentation rather than direct collapsing. However, more precise observation of the kinematics of the surrounding gas on a small scale \citep{vorobyov2021}, complemented with orbital information coming from astrometrical measurements, is required to evaluate this hypothesis. 

In this context, IRAS 04158+2805 is of interest as \cite{Ragusa+2021} reported the detection of two circumstellar disks, similar to HBC 494 in terms of small to moderate misalignment, and an external circumbinary disk. However, in contrast to IRAS 04158+2805, no circumbinary disk was detected in HBC 494. This difference could be explained either by the absorption of the circumbinary disk emission due to cloud contamination, or because HBC 494 had enough time to sufficiently empty the circumbinary mass reservoir. Since most hydrodynamical models of star formation naturally produce young binaries with individual disks and a surrounding circumbinary disk \citep{2018MNRAS.475.5618B, Bate2019, Kuruwita+2020}, it is likely that a circumbinary disk formed at some earlier evolutionary stage of HBC 494. If the circumbinary disk is actually there, but remains undetected, the supply of material could extend the disk lifetime of the circumstellar disks in HBC 494. Deeper multi-wavelength observations would provide more information, enabling us to solve this binary riddle.

\section{Conclusions} 

In this work, we presented high-resolution 1.3 mm observations of HBC 494 with ALMA. The unprecedented angular resolution in this source (0\farcs027), reveals that HBC 494 is a binary that can be resolved into two disks, HBC 494 N and HBC 494 S. The disks have a projected separation of $\sim$75 au. We derived sizes, orientations, inclinations, and dust masses for both components. Both objects appear to be in a quasi-coplanar configuration and with their sizes halted by dynamical evolution, besides preserving high masses, common to FUors and EXors than typical (not eruptive) Class 0/I disks. Comparing the two sources, we noticed that HBC 494 N is $\sim$5 times brighter/more massive and  $\sim$2 times bigger than HBC 494 S. 

The gas kinematics was analyzed at two different spatial scales: 8000 au and 150 au. At large-scale, we have obtained similar results as those presented in \citet[][]{ruizrodriguez2017}, revealing the wide outflow arcs (traced by the $^{12}$CO) and the infalling envelopes ($^{13}$CO and C$^{18}$O). At the small-scale, we detected depleted gas emission (cavities) for both disks  ($^{12}$CO and $^{13}$CO, moment 0). They were probably formed by slow outflows and jets coming from the disks along the line of sight or by optical depth effects and over-subtraction of the continuum emission. The dynamics of both disks in such scales were masked principally by cloud contamination. 

Further observations with similar resolution but of optically thinner molecular lines may lead to the characterization of the dynamical interaction of the two components. For example, the identification of rotational signatures from the disks can be used to identify the dynamical masses of their central stars. With this information, we could also constrain truncation radii. Additionally, observations of small-scale gas structures, allied to astrometric measurements, can be used to constrain the possible formation scenarios.

We conclude that the HBC 494 system constitutes a test bed for eruptive binary-disk interactions and the connection between stellar multiplicity and accretion/luminosity processes (such as outbursts). Also, it provides a statistics enhancement about the rarely observed systems FUors and Exors.

\label{sec:sum}

\section*{Acknowledgements}

We want to thank the referee for their comments and suggestions that significantly improved the manuscript. This paper makes use of the following ALMA data: ADS/JAO.ALMA  2016.1.00630.S. ALMA is a partnership of ESO (representing its member states), NSF (USA) and NINS (Japan), together with NRC (Canada), MOST and ASIAA (Taiwan), and KASI (Republic of Korea), in cooperation with the Republic of Chile. The Joint ALMA Observatory is operated by ESO, AUI/NRAO and NAOJ.  The National Radio Astronomy Observatory is a facility of the National Science Foundation operated under cooperative agreement by Associated Universities, Inc. This work is funded by ANID -- Millennium Science Initiative Program -- Center Code NCN2021\_080. P.H.N. acknowledges support from the Joint China-Chile Committee fund and the ANID Doctorado Nacional grant 21221084 from the government of Chile. A.Z. acknowledges support from the FONDECYT Iniciaci\'on en investigaci\'on project number 11190837. S.P. acknowledges support from FONDECYT Regular project 1191934. L.A.C. acknowledges support from the FONDECYT project number 1211656.  M.M. acknowledges financial support by Fondos de Investigaci\'on 2022 de la Universidad Vi\~na del Mar. J.C. and M.M. acknowledge support from ANID, -- Millennium Science Initiative Program -- NCN19\_171. S.C. acknowledges support from Agencia Nacional de  Investigaci\'on y Desarrollo de Chile (ANID) given by FONDECYT Regular grant 1211496, and ANID project Data Observatory Foundation DO210001. T.B. acknowledges financial support from the Joint Committee ESO-Government of Chile fund and the FONDECYT postdoctorado project number 3230470. This project has received funding from the European Union (ERC, Stellar-MADE, No. 101042275).
\section*{Data Availability}


 This paper makes use of the following ALMA data: ADS/JAO.ALMA 2016.1.00630.S. The data are downloadable from the official ALMA archive and public. ALMA is a partnership of ESO (representing its member states), NSF (USA) and NINS (Japan), together with NRC (Canada), MOST and ASIAA (Taiwan), and KASI (Republic of Korea), in cooperation with the Republic of Chile. The Joint ALMA Observatory is operated by ESO, AUI/NRAO and NAOJ.



\bibliographystyle{mnras}
\bibliography{mnras_template} 

\begin{thebibliography}{}
\makeatletter
\relax
\def\mn@urlcharsother{\let\do\@makeother \do\$\do\&\do\#\do\^\do\_\do\%\do\~}
\def\mn@doi{\begingroup\mn@urlcharsother \@ifnextchar [ {\mn@doi@}
  {\mn@doi@[]}}
\def\mn@doi@[#1]#2{\def\@tempa{#1}\ifx\@tempa\@empty \href
  {http://dx.doi.org/#2} {doi:#2}\else \href {http://dx.doi.org/#2} {#1}\fi
  \endgroup}
\def\mn@eprint#1#2{\mn@eprint@#1:#2::\@nil}
\def\mn@eprint@arXiv#1{\href {http://arxiv.org/abs/#1} {{\tt arXiv:#1}}}
\def\mn@eprint@dblp#1{\href {http://dblp.uni-trier.de/rec/bibtex/#1.xml}
  {dblp:#1}}
\def\mn@eprint@#1:#2:#3:#4\@nil{\def\@tempa {#1}\def\@tempb {#2}\def\@tempc
  {#3}\ifx \@tempc \@empty \let \@tempc \@tempb \let \@tempb \@tempa \fi \ifx
  \@tempb \@empty \def\@tempb {arXiv}\fi \@ifundefined
  {mn@eprint@\@tempb}{\@tempb:\@tempc}{\expandafter \expandafter \csname
  mn@eprint@\@tempb\endcsname \expandafter{\@tempc}}}

\bibitem[\protect\citeauthoryear{{Andrews} \& {Williams}}{{Andrews} \&
  {Williams}}{2005}]{andrews2005}
{Andrews} S.~M.,  {Williams} J.~P.,  2005, \mn@doi [\apj] {10.1086/432712},
  \href {http://adsabs.harvard.edu/abs/2005ApJ...631.1134A} {631, 1134}

\bibitem[\protect\citeauthoryear{{Arce} \& {Sargent}}{{Arce} \&
  {Sargent}}{2006}]{arce2006}
{Arce} H.~G.,  {Sargent} A.~I.,  2006, \mn@doi [\apj] {10.1086/505104}, \href
  {https://ui.adsabs.harvard.edu/abs/2006ApJ...646.1070A} {646, 1070}

\bibitem[\protect\citeauthoryear{{Armitage}, {Livio}  \& {Pringle}}{{Armitage}
  et~al.}{2001}]{armitage2001}
{Armitage} P.~J.,  {Livio} M.,   {Pringle} J.~E.,  2001, \mn@doi [\mnras]
  {10.1046/j.1365-8711.2001.04356.x}, \href
  {http://adsabs.harvard.edu/abs/2001MNRAS.324..705A} {324, 705}

\bibitem[\protect\citeauthoryear{{Audard} et~al.,}{{Audard}
  et~al.}{2014}]{audard2014}
{Audard} M.,  et~al., 2014, \mn@doi [Protostars and Planets VI]
  {10.2458/azu_uapress_9780816531240-ch017}, \href
  {http://adsabs.harvard.edu/abs/2014prpl.conf..387A} {pp 387--410}

\bibitem[\protect\citeauthoryear{{Bate}}{{Bate}}{2000}]{2000MNRAS.314...33B}
{Bate} M.~R.,  2000, \mn@doi [\mnras] {10.1046/j.1365-8711.2000.03333.x}, \href
  {https://ui.adsabs.harvard.edu/abs/2000MNRAS.314...33B} {314, 33}

\bibitem[\protect\citeauthoryear{{Bate}}{{Bate}}{2018}]{2018MNRAS.475.5618B}
{Bate} M.~R.,  2018, \mn@doi [\mnras] {10.1093/mnras/sty169}, \href
  {https://ui.adsabs.harvard.edu/abs/2018MNRAS.475.5618B} {475, 5618}

\bibitem[\protect\citeauthoryear{{Bate}}{{Bate}}{2019}]{Bate2019}
{Bate} M.~R.,  2019, \mn@doi [\mnras] {10.1093/mnras/stz103}, \href
  {https://ui.adsabs.harvard.edu/abs/2019MNRAS.484.2341B} {484, 2341}

\bibitem[\protect\citeauthoryear{{Bate} \& {Burkert}}{{Bate} \&
  {Burkert}}{1997}]{1997MNRAS.288.1060B}
{Bate} M.~R.,  {Burkert} A.,  1997, \mn@doi [\mnras]
  {10.1093/mnras/288.4.1060}, \href
  {https://ui.adsabs.harvard.edu/abs/1997MNRAS.288.1060B} {288, 1060}

\bibitem[\protect\citeauthoryear{{Beckwith}, {Sargent}, {Chini}  \&
  {Guesten}}{{Beckwith} et~al.}{1990}]{beckwith1990}
{Beckwith} S. V.~W.,  {Sargent} A.~I.,  {Chini} R.~S.,   {Guesten} R.,  1990,
  \mn@doi [\aj] {10.1086/115385}, \href
  {https://ui.adsabs.harvard.edu/abs/1990AJ.....99..924B} {99, 924}

\bibitem[\protect\citeauthoryear{{Boehler}, {Weaver}, {Isella}, {Ricci},
  {Grady}, {Carpenter}  \& {Perez}}{{Boehler} et~al.}{2017}]{boehler2017}
{Boehler} Y.,  {Weaver} E.,  {Isella} A.,  {Ricci} L.,  {Grady} C.,
  {Carpenter} J.,   {Perez} L.,  2017, \mn@doi [\apj]
  {10.3847/1538-4357/aa696c}, \href
  {https://ui.adsabs.harvard.edu/abs/2017ApJ...840...60B} {840, 60}

\bibitem[\protect\citeauthoryear{{Bonnell}}{{Bonnell}}{1994}]{1994MNRAS.269..837B}
{Bonnell} I.~A.,  1994, \mn@doi [\mnras] {10.1093/mnras/269.3.837}, \href
  {https://ui.adsabs.harvard.edu/abs/1994MNRAS.269..837B} {269, 837}

\bibitem[\protect\citeauthoryear{{Bonnell} \& {Bastien}}{{Bonnell} \&
  {Bastien}}{1992}]{bonnell1992}
{Bonnell} I.,  {Bastien} P.,  1992, \mn@doi [\apjl] {10.1086/186663}, \href
  {http://adsabs.harvard.edu/abs/1992ApJ...401L..31B} {401, L31}

\bibitem[\protect\citeauthoryear{{Bonnell} \& {Bate}}{{Bonnell} \&
  {Bate}}{1994a}]{1994MNRAS.269L..45B}
{Bonnell} I.~A.,  {Bate} M.~R.,  1994a, \mn@doi [\mnras]
  {10.1093/mnras/269.1.L45}, \href
  {https://ui.adsabs.harvard.edu/abs/1994MNRAS.269L..45B} {269, L45}

\bibitem[\protect\citeauthoryear{{Bonnell} \& {Bate}}{{Bonnell} \&
  {Bate}}{1994b}]{1994MNRAS.271..999B}
{Bonnell} I.~A.,  {Bate} M.~R.,  1994b, \mn@doi [\mnras]
  {10.1093/mnras/271.4.999}, \href
  {https://ui.adsabs.harvard.edu/abs/1994MNRAS.271..999B} {271, 999}

\bibitem[\protect\citeauthoryear{{Borchert}, {Price}, {Pinte}  \&
  {Cuello}}{{Borchert} et~al.}{2022a}]{Borchert+2022a}
{Borchert} E. M.~A.,  {Price} D.~J.,  {Pinte} C.,   {Cuello} N.,  2022a,
  \mn@doi [\mnras] {10.1093/mnrasl/slab123}, \href
  {https://ui.adsabs.harvard.edu/abs/2022MNRAS.510L..37B} {510, L37}

\bibitem[\protect\citeauthoryear{{Borchert}, {Price}, {Pinte}  \&
  {Cuello}}{{Borchert} et~al.}{2022b}]{borchert2022}
{Borchert} E. M.~A.,  {Price} D.~J.,  {Pinte} C.,   {Cuello} N.,  2022b,
  \mn@doi [\mnras] {10.1093/mnras/stac2872}, \href
  {https://ui.adsabs.harvard.edu/abs/2022MNRAS.517.4436B} {517, 4436}

\bibitem[\protect\citeauthoryear{{Borchert}, {Price}, {Pinte}  \&
  {Cuello}}{{Borchert} et~al.}{2022c}]{Borchert+2022b}
{Borchert} E. M.~A.,  {Price} D.~J.,  {Pinte} C.,   {Cuello} N.,  2022c,
  \mn@doi [\mnras] {10.1093/mnras/stac2872}, \href
  {https://ui.adsabs.harvard.edu/abs/2022MNRAS.517.4436B} {517, 4436}

\bibitem[\protect\citeauthoryear{{Boss} \& {Bodenheimer}}{{Boss} \&
  {Bodenheimer}}{1979}]{1979ApJ...234..289B}
{Boss} A.~P.,  {Bodenheimer} P.,  1979, \mn@doi [\apj] {10.1086/157497}, \href
  {https://ui.adsabs.harvard.edu/abs/1979ApJ...234..289B} {234, 289}

\bibitem[\protect\citeauthoryear{{Burkert}, {Bate}  \& {Bodenheimer}}{{Burkert}
  et~al.}{1997}]{1997MNRAS.289..497B}
{Burkert} A.,  {Bate} M.~R.,   {Bodenheimer} P.,  1997, \mn@doi [\mnras]
  {10.1093/mnras/289.3.497}, \href
  {https://ui.adsabs.harvard.edu/abs/1997MNRAS.289..497B} {289, 497}

\bibitem[\protect\citeauthoryear{{CASA Team} et~al.,}{{CASA Team}
  et~al.}{2022}]{2022PASP..134k4501C}
{CASA Team} et~al., 2022, \mn@doi [\pasp] {10.1088/1538-3873/ac9642}, \href
  {https://ui.adsabs.harvard.edu/abs/2022PASP..134k4501C} {134, 114501}

\bibitem[\protect\citeauthoryear{{Chiang}, {Reipurth}, {Walawender},
  {Connelley}, {Pessev}, {Geballe}, {Best}  \& {Paegert}}{{Chiang}
  et~al.}{2015}]{chiang2015}
{Chiang} H.-F.,  {Reipurth} B.,  {Walawender} J.,  {Connelley} M.~S.,  {Pessev}
  P.,  {Geballe} T.~R.,  {Best} W. M.~J.,   {Paegert} M.,  2015, \mn@doi [\apj]
  {10.1088/0004-637X/805/1/54}, \href
  {https://ui.adsabs.harvard.edu/abs/2015ApJ...805...54C} {805, 54}

\bibitem[\protect\citeauthoryear{{Cieza} et~al.,}{{Cieza}
  et~al.}{2018}]{cieza2018}
{Cieza} L.~A.,  et~al., 2018, \mn@doi [\mnras] {10.1093/mnras/stx3059}, \href
  {https://ui.adsabs.harvard.edu/abs/2018MNRAS.474.4347C} {474, 4347}

\bibitem[\protect\citeauthoryear{{Clarke} \& {Pringle}}{{Clarke} \&
  {Pringle}}{1993}]{ClarkePringle1993}
{Clarke} C.~J.,  {Pringle} J.~E.,  1993, \mn@doi [\mnras]
  {10.1093/mnras/261.1.190}, \href
  {https://ui.adsabs.harvard.edu/abs/1993MNRAS.261..190C} {261, 190}

\bibitem[\protect\citeauthoryear{{Connelley} \& {Reipurth}}{{Connelley} \&
  {Reipurth}}{2018}]{connelley2018}
{Connelley} M.~S.,  {Reipurth} B.,  2018, \mn@doi [\apj]
  {10.3847/1538-4357/aaba7b}, \href
  {https://ui.adsabs.harvard.edu/abs/2018ApJ...861..145C} {861, 145}

\bibitem[\protect\citeauthoryear{{Cruz-S{\'a}enz de Miera}, {K{\'o}sp{\'a}l},
  {{\'A}brah{\'a}m}, {Liu}  \& {Takami}}{{Cruz-S{\'a}enz de Miera}
  et~al.}{2019}]{cruz2019}
{Cruz-S{\'a}enz de Miera} F.,  {K{\'o}sp{\'a}l} {\'A}.,  {{\'A}brah{\'a}m} P.,
  {Liu} H.~B.,   {Takami} M.,  2019, \mn@doi [\apjl]
  {10.3847/2041-8213/ab39ea}, \href
  {https://ui.adsabs.harvard.edu/abs/2019ApJ...882L...4C} {882, L4}

\bibitem[\protect\citeauthoryear{{Cuello} et~al.,}{{Cuello}
  et~al.}{2019}]{2019MNRAS.483.4114C}
{Cuello} N.,  et~al., 2019, \mn@doi [\mnras]
  {10.1093/mnras/sty332510.48550/arXiv.1812.00961}, \href
  {https://ui.adsabs.harvard.edu/abs/2019MNRAS.483.4114C} {483, 4114}

\bibitem[\protect\citeauthoryear{{Cuello} et~al.,}{{Cuello}
  et~al.}{2020}]{cuello2020}
{Cuello} N.,  et~al., 2020, \mn@doi [\mnras] {10.1093/mnras/stz2938}, \href
  {https://ui.adsabs.harvard.edu/abs/2020MNRAS.491..504C} {491, 504}

\bibitem[\protect\citeauthoryear{{Cuello}, {M{\'e}nard}  \& {Price}}{{Cuello}
  et~al.}{2023}]{Cuello+2023}
{Cuello} N.,  {M{\'e}nard} F.,   {Price} D.~J.,  2023, \mn@doi [European
  Physical Journal Plus] {10.1140/epjp/s13360-022-03602-w}, \href
  {https://ui.adsabs.harvard.edu/abs/2023EPJP..138...11C} {138, 11}

\bibitem[\protect\citeauthoryear{{Dullemond}, {K{\"u}ffmeier}, {Goicovic},
  {Fukagawa}, {Oehl}  \& {Kramer}}{{Dullemond} et~al.}{2019}]{Dullemond+2019}
{Dullemond} C.~P.,  {K{\"u}ffmeier} M.,  {Goicovic} F.,  {Fukagawa} M.,  {Oehl}
  V.,   {Kramer} M.,  2019, \mn@doi [\aap] {10.1051/0004-6361/201832632}, \href
  {https://ui.adsabs.harvard.edu/abs/2019A&A...628A..20D} {628, A20}

\bibitem[\protect\citeauthoryear{{Dunham} \& {Vorobyov}}{{Dunham} \&
  {Vorobyov}}{2012}]{dunham2012}
{Dunham} M.~M.,  {Vorobyov} E.~I.,  2012, \mn@doi [\apj]
  {10.1088/0004-637X/747/1/52}, \href
  {https://ui.adsabs.harvard.edu/abs/2012ApJ...747...52D} {747, 52}

\bibitem[\protect\citeauthoryear{{Dunhill}, {Cuadra}  \& {Dougados}}{{Dunhill}
  et~al.}{2015}]{2015MNRAS.448.3545D}
{Dunhill} A.~C.,  {Cuadra} J.,   {Dougados} C.,  2015, \mn@doi [\mnras]
  {10.1093/mnras/stv284}, \href
  {https://ui.adsabs.harvard.edu/abs/2015MNRAS.448.3545D} {448, 3545}

\bibitem[\protect\citeauthoryear{{Dutrey}, {Guilloteau}  \& {Simon}}{{Dutrey}
  et~al.}{1994}]{1994A&A...286..149D}
{Dutrey} A.,  {Guilloteau} S.,   {Simon} M.,  1994, \aap, \href
  {https://ui.adsabs.harvard.edu/abs/1994A&A...286..149D} {286, 149}

\bibitem[\protect\citeauthoryear{{Evans} Neal~J. et~al.,}{{Evans}
  et~al.}{2009}]{evans2009}
{Evans} Neal~J. I.,  et~al., 2009, \mn@doi [\apjs]
  {10.1088/0067-0049/181/2/321}, \href
  {https://ui.adsabs.harvard.edu/abs/2009ApJS..181..321E} {181, 321}

\bibitem[\protect\citeauthoryear{{Frank} et~al.,}{{Frank}
  et~al.}{2014}]{frank2014}
{Frank} A.,  et~al., 2014, in {Beuther} H.,  {Klessen} R.~S.,  {Dullemond}
  C.~P.,   {Henning} T.,  eds, Protostars and Planets VI. p.~451 (\mn@eprint
  {arXiv} {1402.3553}), \mn@doi{10.2458/azu_uapress_9780816531240-ch020}

\bibitem[\protect\citeauthoryear{{Giannini} et~al.,}{{Giannini}
  et~al.}{2022}]{2022ApJ...929..129G}
{Giannini} T.,  et~al., 2022, \mn@doi [\apj] {10.3847/1538-4357/ac5a49}, \href
  {https://ui.adsabs.harvard.edu/abs/2022ApJ...929..129G} {929, 129}

\bibitem[\protect\citeauthoryear{{Hales} et~al.,}{{Hales}
  et~al.}{2018}]{hales2018}
{Hales} A.~S.,  et~al., 2018, \mn@doi [\apj] {10.3847/1538-4357/aac018}, \href
  {https://ui.adsabs.harvard.edu/abs/2018ApJ...859..111H} {859, 111}

\bibitem[\protect\citeauthoryear{{Hales} et~al.,}{{Hales}
  et~al.}{2020}]{hales2020}
{Hales} A.~S.,  et~al., 2020, \mn@doi [\apj] {10.3847/1538-4357/aba3c4}, \href
  {https://ui.adsabs.harvard.edu/abs/2020ApJ...900....7H} {900, 7}

\bibitem[\protect\citeauthoryear{{Harris}, {Andrews}, {Wilner}  \&
  {Kraus}}{{Harris} et~al.}{2012}]{harris2012}
{Harris} R.~J.,  {Andrews} S.~M.,  {Wilner} D.~J.,   {Kraus} A.~L.,  2012,
  \mn@doi [\apj] {10.1088/0004-637X/751/2/115}, \href
  {https://ui.adsabs.harvard.edu/abs/2012ApJ...751..115H} {751, 115}

\bibitem[\protect\citeauthoryear{{Herbig}}{{Herbig}}{1966}]{herbig1966}
{Herbig} G.~H.,  1966, \mn@doi [Vistas in Astronomy]
  {10.1016/0083-6656(66)90025-0}, \href
  {http://adsabs.harvard.edu/abs/1966VA......8..109H} {8, 109}

\bibitem[\protect\citeauthoryear{{Herbig}}{{Herbig}}{2007}]{herbig2007}
{Herbig} G.~H.,  2007, \mn@doi [\aj] {10.1086/517494}, \href
  {https://ui.adsabs.harvard.edu/abs/2007AJ....133.2679H} {133, 2679}

\bibitem[\protect\citeauthoryear{{Jurdana-{\v{S}}epi{\'c}}, {Munari},
  {Antoniucci}, {Giannini}  \& {Lorenzetti}}{{Jurdana-{\v{S}}epi{\'c}}
  et~al.}{2018}]{2018A&A...614A...9J}
{Jurdana-{\v{S}}epi{\'c}} R.,  {Munari} U.,  {Antoniucci} S.,  {Giannini} T.,
  {Lorenzetti} D.,  2018, \mn@doi [\aap] {10.1051/0004-6361/201732131}, \href
  {https://ui.adsabs.harvard.edu/abs/2018A&A...614A...9J} {614, A9}

\bibitem[\protect\citeauthoryear{{Kenyon}, {Hartmann}, {Strom}  \&
  {Strom}}{{Kenyon} et~al.}{1990}]{kenyon1990}
{Kenyon} S.~J.,  {Hartmann} L.~W.,  {Strom} K.~M.,   {Strom} S.~E.,  1990,
  \mn@doi [\aj] {10.1086/115380}, \href
  {https://ui.adsabs.harvard.edu/abs/1990AJ.....99..869K} {99, 869}

\bibitem[\protect\citeauthoryear{{K{\'o}sp{\'a}l} et~al.,}{{K{\'o}sp{\'a}l}
  et~al.}{2017}]{kospal2017}
{K{\'o}sp{\'a}l} {\'A}.,  et~al., 2017, \mn@doi [\apj]
  {10.3847/1538-4357/aa7683}, \href
  {https://ui.adsabs.harvard.edu/abs/2017ApJ...843...45K} {843, 45}

\bibitem[\protect\citeauthoryear{{Kratter} \& {Lodato}}{{Kratter} \&
  {Lodato}}{2016}]{kratter2016}
{Kratter} K.,  {Lodato} G.,  2016, \mn@doi [\araa]
  {10.1146/annurev-astro-081915-023307}, \href
  {https://ui.adsabs.harvard.edu/abs/2016ARA&A..54..271K} {54, 271}

\bibitem[\protect\citeauthoryear{{Kraus}, {Ireland}, {Martinache}  \&
  {Hillenbrand}}{{Kraus} et~al.}{2011}]{kraus2011}
{Kraus} A.~L.,  {Ireland} M.~J.,  {Martinache} F.,   {Hillenbrand} L.~A.,
  2011, \mn@doi [\apj] {10.1088/0004-637X/731/1/8}, \href
  {https://ui.adsabs.harvard.edu/abs/2011ApJ...731....8K} {731, 8}

\bibitem[\protect\citeauthoryear{{Kraus}, {Ireland}, {Hillenbrand}  \&
  {Martinache}}{{Kraus} et~al.}{2012}]{kraus2012}
{Kraus} A.~L.,  {Ireland} M.~J.,  {Hillenbrand} L.~A.,   {Martinache} F.,
  2012, \mn@doi [\apj] {10.1088/0004-637X/745/1/19}, \href
  {https://ui.adsabs.harvard.edu/abs/2012ApJ...745...19K} {745, 19}

\bibitem[\protect\citeauthoryear{{Kuffmeier}, {Goicovic}  \&
  {Dullemond}}{{Kuffmeier} et~al.}{2020}]{Kuffmeier+2020}
{Kuffmeier} M.,  {Goicovic} F.~G.,   {Dullemond} C.~P.,  2020, \mn@doi [\aap]
  {10.1051/0004-6361/201936820}, \href
  {https://ui.adsabs.harvard.edu/abs/2020A&A...633A...3K} {633, A3}

\bibitem[\protect\citeauthoryear{{Kuffmeier}, {Dullemond}, {Reissl}  \&
  {Goicovic}}{{Kuffmeier} et~al.}{2021}]{Kuffmeier+2021}
{Kuffmeier} M.,  {Dullemond} C.~P.,  {Reissl} S.,   {Goicovic} F.~G.,  2021,
  \mn@doi [\aap] {10.1051/0004-6361/202039614}, \href
  {https://ui.adsabs.harvard.edu/abs/2021A&A...656A.161K} {656, A161}

\bibitem[\protect\citeauthoryear{{Kuruwita}, {Federrath}  \&
  {Haugb{\o}lle}}{{Kuruwita} et~al.}{2020}]{Kuruwita+2020}
{Kuruwita} R.~L.,  {Federrath} C.,   {Haugb{\o}lle} T.,  2020, \mn@doi [\aap]
  {10.1051/0004-6361/202038181}, \href
  {https://ui.adsabs.harvard.edu/abs/2020A&A...641A..59K} {641, A59}

\bibitem[\protect\citeauthoryear{{Lai} \& {Mu{\~n}oz}}{{Lai} \&
  {Mu{\~n}oz}}{2022}]{2022arXiv221100028L}
{Lai} D.,  {Mu{\~n}oz} D.~J.,  2022, arXiv e-prints, \href
  {https://ui.adsabs.harvard.edu/abs/2022arXiv221100028L} {p. arXiv:2211.00028}

\bibitem[\protect\citeauthoryear{{Larson}}{{Larson}}{1972}]{larson1972}
{Larson} R.~B.,  1972, \mn@doi [\mnras] {10.1093/mnras/156.4.437}, \href
  {https://ui.adsabs.harvard.edu/abs/1972MNRAS.156..437L} {156, 437}

\bibitem[\protect\citeauthoryear{{Lee}, {Offner}, {Kratter}, {Smullen}  \&
  {Li}}{{Lee} et~al.}{2019}]{2019ApJ...887..232L}
{Lee} A.~T.,  {Offner} S. S.~R.,  {Kratter} K.~M.,  {Smullen} R.~A.,   {Li}
  P.~S.,  2019, \mn@doi [\apj] {10.3847/1538-4357/ab584b}, \href
  {https://ui.adsabs.harvard.edu/abs/2019ApJ...887..232L} {887, 232}

\bibitem[\protect\citeauthoryear{{Lodato} \& {Clarke}}{{Lodato} \&
  {Clarke}}{2004}]{lodato2004}
{Lodato} G.,  {Clarke} C.~J.,  2004, \mn@doi [\mnras]
  {10.1111/j.1365-2966.2004.08112.x}, \href
  {http://adsabs.harvard.edu/abs/2004MNRAS.353..841L} {353, 841}

\bibitem[\protect\citeauthoryear{{Mathieu}, {Stassun}, {Basri}, {Jensen},
  {Johns-Krull}, {Valenti}  \& {Hartmann}}{{Mathieu}
  et~al.}{1997}]{1997AJ....113.1841M}
{Mathieu} R.~D.,  {Stassun} K.,  {Basri} G.,  {Jensen} E. L.~N.,  {Johns-Krull}
  C.~M.,  {Valenti} J.~A.,   {Hartmann} L.~W.,  1997, \mn@doi [\aj]
  {10.1086/118395}, \href
  {https://ui.adsabs.harvard.edu/abs/1997AJ....113.1841M} {113, 1841}

\bibitem[\protect\citeauthoryear{{Maureira}, {Pineda}, {Segura-Cox}, {Caselli},
  {Testi}, {Lodato}, {Loinard}  \& {Hern{\'a}ndez-G{\'o}mez}}{{Maureira}
  et~al.}{2020}]{2020ApJ...897...59M}
{Maureira} M.~J.,  {Pineda} J.~E.,  {Segura-Cox} D.~M.,  {Caselli} P.,  {Testi}
  L.,  {Lodato} G.,  {Loinard} L.,   {Hern{\'a}ndez-G{\'o}mez} A.,  2020,
  \mn@doi [\apj] {10.3847/1538-4357/ab960b}, \href
  {https://ui.adsabs.harvard.edu/abs/2020ApJ...897...59M} {897, 59}

\bibitem[\protect\citeauthoryear{{McMullin}, {Waters}, {Schiebel}, {Young}  \&
  {Golap}}{{McMullin} et~al.}{2007}]{2007ASPC..376..127M}
{McMullin} J.~P.,  {Waters} B.,  {Schiebel} D.,  {Young} W.,   {Golap} K.,
  2007, in {Shaw} R.~A.,  {Hill} F.,   {Bell} D.~J.,  eds,  Astronomical
  Society of the Pacific Conference Series Vol. 376, Astronomical Data Analysis
  Software and Systems XVI. p.~127

\bibitem[\protect\citeauthoryear{{Menten}, {Reid}, {Forbrich}  \&
  {Brunthaler}}{{Menten} et~al.}{2007}]{2007A&A...474..515M}
{Menten} K.~M.,  {Reid} M.~J.,  {Forbrich} J.,   {Brunthaler} A.,  2007,
  \mn@doi [\aap] {10.1051/0004-6361:20078247}, \href
  {https://ui.adsabs.harvard.edu/abs/2007A&A...474..515M} {474, 515}

\bibitem[\protect\citeauthoryear{{Nealon}, {Cuello}  \& {Alexander}}{{Nealon}
  et~al.}{2020}]{Nealon+2020}
{Nealon} R.,  {Cuello} N.,   {Alexander} R.,  2020, \mn@doi [\mnras]
  {10.1093/mnras/stz3186}, \href
  {https://ui.adsabs.harvard.edu/abs/2020MNRAS.491.4108N} {491, 4108}

\bibitem[\protect\citeauthoryear{{Offner}, {Dunham}, {Lee}, {Arce}  \&
  {Fielding}}{{Offner} et~al.}{2016}]{2016ApJ...827L..11O}
{Offner} S. S.~R.,  {Dunham} M.~M.,  {Lee} K.~I.,  {Arce} H.~G.,   {Fielding}
  D.~B.,  2016, \mn@doi [\apjl] {10.3847/2041-8205/827/1/L11}, \href
  {https://ui.adsabs.harvard.edu/abs/2016ApJ...827L..11O} {827, L11}

\bibitem[\protect\citeauthoryear{{Offner}, {Moe}, {Kratter}, {Sadavoy},
  {Jensen}  \& {Tobin}}{{Offner} et~al.}{2022}]{offner2022}
{Offner} S. S.~R.,  {Moe} M.,  {Kratter} K.~M.,  {Sadavoy} S.~I.,  {Jensen} E.
  L.~N.,   {Tobin} J.~J.,  2022, arXiv e-prints, \href
  {https://ui.adsabs.harvard.edu/abs/2022arXiv220310066O} {p. arXiv:2203.10066}

\bibitem[\protect\citeauthoryear{{P{\'e}rez} et~al.,}{{P{\'e}rez}
  et~al.}{2020}]{perez2020}
{P{\'e}rez} S.,  et~al., 2020, \mn@doi [\apj] {10.3847/1538-4357/ab5c1b}, \href
  {https://ui.adsabs.harvard.edu/abs/2020ApJ...889...59P} {889, 59}

\bibitem[\protect\citeauthoryear{{Postel}, {Audard}, {Vorobyov}, {Dionatos},
  {Rab}  \& {G{\"u}del}}{{Postel} et~al.}{2019}]{2019A&A...631A..30P}
{Postel} A.,  {Audard} M.,  {Vorobyov} E.,  {Dionatos} O.,  {Rab} C.,
  {G{\"u}del} M.,  2019, \mn@doi [\aap] {10.1051/0004-6361/201935601}, \href
  {https://ui.adsabs.harvard.edu/abs/2019A&A...631A..30P} {631, A30}

\bibitem[\protect\citeauthoryear{{Pueyo} et~al.,}{{Pueyo}
  et~al.}{2012}]{pueyo2012}
{Pueyo} L.,  et~al., 2012, \mn@doi [\apj] {10.1088/0004-637X/757/1/57}, \href
  {https://ui.adsabs.harvard.edu/abs/2012ApJ...757...57P} {757, 57}

\bibitem[\protect\citeauthoryear{{Ragusa} et~al.,}{{Ragusa}
  et~al.}{2021}]{Ragusa+2021}
{Ragusa} E.,  et~al., 2021, \mn@doi [\mnras] {10.1093/mnras/stab2179}, \href
  {https://ui.adsabs.harvard.edu/abs/2021MNRAS.507.1157R} {507, 1157}

\bibitem[\protect\citeauthoryear{{Reipurth}}{{Reipurth}}{1985}]{reipurth1985}
{Reipurth} B.,  1985, \aaps, \href
  {https://ui.adsabs.harvard.edu/abs/1985A&AS...61..319R} {61, 319}

\bibitem[\protect\citeauthoryear{{Reipurth} \& {Aspin}}{{Reipurth} \&
  {Aspin}}{2004}]{Reipurth+2004}
{Reipurth} B.,  {Aspin} C.,  2004, \mn@doi [\apjl] {10.1086/422250}, \href
  {https://ui.adsabs.harvard.edu/abs/2004ApJ...608L..65R} {608, L65}

\bibitem[\protect\citeauthoryear{{Reipurth} \& {Bally}}{{Reipurth} \&
  {Bally}}{1986}]{reipurth1986}
{Reipurth} B.,  {Bally} J.,  1986, \mn@doi [\nat] {10.1038/320336a0}, \href
  {https://ui.adsabs.harvard.edu/abs/1986Natur.320..336R} {320, 336}

\bibitem[\protect\citeauthoryear{{Rosotti} \& {Clarke}}{{Rosotti} \&
  {Clarke}}{2018}]{rosotti2018}
{Rosotti} G.~P.,  {Clarke} C.~J.,  2018, \mn@doi [\mnras]
  {10.1093/mnras/stx2769}, \href
  {https://ui.adsabs.harvard.edu/abs/2018MNRAS.473.5630R} {473, 5630}

\bibitem[\protect\citeauthoryear{{Ru{\'\i}z-Rodr{\'\i}guez}
  et~al.,}{{Ru{\'\i}z-Rodr{\'\i}guez} et~al.}{2017}]{ruizrodriguez2017}
{Ru{\'\i}z-Rodr{\'\i}guez} D.,  et~al., 2017, \mn@doi [\mnras]
  {10.1093/mnras/stw3378}, \href
  {https://ui.adsabs.harvard.edu/abs/2017MNRAS.466.3519R} {466, 3519}

\bibitem[\protect\citeauthoryear{{Ru{\'\i}z-Rodr{\'\i}guez}, {Williams},
  {Kastner}, {Cieza}, {Leemker}  \& {Principe}}{{Ru{\'\i}z-Rodr{\'\i}guez}
  et~al.}{2022}]{ruizrodriguez2022}
{Ru{\'\i}z-Rodr{\'\i}guez} D.~A.,  {Williams} J.~P.,  {Kastner} J.~H.,  {Cieza}
  L.,  {Leemker} M.,   {Principe} D.~A.,  2022, \mn@doi [\mnras]
  {10.1093/mnras/stac1879}, \href
  {https://ui.adsabs.harvard.edu/abs/2022MNRAS.515.2646R} {515, 2646}

\bibitem[\protect\citeauthoryear{{Takami} et~al.,}{{Takami}
  et~al.}{2019}]{takami2019}
{Takami} M.,  et~al., 2019, \mn@doi [\apj] {10.3847/1538-4357/ab43c8}, \href
  {https://ui.adsabs.harvard.edu/abs/2019ApJ...884..146T} {884, 146}

\bibitem[\protect\citeauthoryear{{Tobin} et~al.,}{{Tobin}
  et~al.}{2016a}]{2016Natur.538..483T}
{Tobin} J.~J.,  et~al., 2016a, \mn@doi [\nat] {10.1038/nature20094}, \href
  {https://ui.adsabs.harvard.edu/abs/2016Natur.538..483T} {538, 483}

\bibitem[\protect\citeauthoryear{{Tobin} et~al.,}{{Tobin}
  et~al.}{2016b}]{2016ApJ...818...73T}
{Tobin} J.~J.,  et~al., 2016b, \mn@doi [\apj] {10.3847/0004-637X/818/1/73},
  \href {https://ui.adsabs.harvard.edu/abs/2016ApJ...818...73T} {818, 73}

\bibitem[\protect\citeauthoryear{{Tobin} et~al.,}{{Tobin}
  et~al.}{2020}]{tobin2020}
{Tobin} J.~J.,  et~al., 2020, \mn@doi [\apj] {10.3847/1538-4357/ab6f64}, \href
  {https://ui.adsabs.harvard.edu/abs/2020ApJ...890..130T} {890, 130}

\bibitem[\protect\citeauthoryear{{Tobin} et~al.,}{{Tobin}
  et~al.}{2022}]{tobin2022}
{Tobin} J.~J.,  et~al., 2022, \mn@doi [\apj] {10.3847/1538-4357/ac36d2}, \href
  {https://ui.adsabs.harvard.edu/abs/2022ApJ...925...39T} {925, 39}

\bibitem[\protect\citeauthoryear{{Vorobyov} \& {Basu}}{{Vorobyov} \&
  {Basu}}{2005}]{vorobyov2005}
{Vorobyov} E.~I.,  {Basu} S.,  2005, \mn@doi [\apjl] {10.1086/498303}, \href
  {http://adsabs.harvard.edu/abs/2005ApJ...633L.137V} {633, L137}

\bibitem[\protect\citeauthoryear{{Vorobyov} \& {Basu}}{{Vorobyov} \&
  {Basu}}{2015}]{vorobyov2015}
{Vorobyov} E.~I.,  {Basu} S.,  2015, \mn@doi [\apj]
  {10.1088/0004-637X/805/2/115}, \href
  {https://ui.adsabs.harvard.edu/abs/2015ApJ...805..115V} {805, 115}

\bibitem[\protect\citeauthoryear{{Vorobyov}, {Elbakyan}, {Liu}  \&
  {Takami}}{{Vorobyov} et~al.}{2021}]{vorobyov2021}
{Vorobyov} E.~I.,  {Elbakyan} V.~G.,  {Liu} H.~B.,   {Takami} M.,  2021,
  \mn@doi [\aap] {10.1051/0004-6361/202039391}, \href
  {https://ui.adsabs.harvard.edu/abs/2021A&A...647A..44V} {647, A44}

\bibitem[\protect\citeauthoryear{{Weaver}, {Isella}  \& {Boehler}}{{Weaver}
  et~al.}{2018}]{weaver2018}
{Weaver} E.,  {Isella} A.,   {Boehler} Y.,  2018, \mn@doi [\apj]
  {10.3847/1538-4357/aaa481}, \href
  {https://ui.adsabs.harvard.edu/abs/2018ApJ...853..113W} {853, 113}

\bibitem[\protect\citeauthoryear{{Zhu}, {Hartmann}, {Gammie}  \&
  {McKinney}}{{Zhu} et~al.}{2009}]{zhu2009}
{Zhu} Z.,  {Hartmann} L.,  {Gammie} C.,   {McKinney} J.~C.,  2009, \mn@doi
  [\apj] {10.1088/0004-637X/701/1/620}, \href
  {http://adsabs.harvard.edu/abs/2009ApJ...701..620Z} {701, 620}

\bibitem[\protect\citeauthoryear{{Zhu}, {Hartmann}, {Nelson}  \&
  {Gammie}}{{Zhu} et~al.}{2012}]{zhu2012}
{Zhu} Z.,  {Hartmann} L.,  {Nelson} R.~P.,   {Gammie} C.~F.,  2012, \mn@doi
  [\apj] {10.1088/0004-637X/746/1/110}, \href
  {http://adsabs.harvard.edu/abs/2012ApJ...746..110Z} {746, 110}

\bibitem[\protect\citeauthoryear{{Zurlo} et~al.,}{{Zurlo}
  et~al.}{2020a}]{zurlo2020b}
{Zurlo} A.,  et~al., 2020a, \mn@doi [\mnras] {10.1093/mnras/staa1886}, \href
  {https://ui.adsabs.harvard.edu/abs/2020MNRAS.496.5089Z} {496, 5089}

\bibitem[\protect\citeauthoryear{{Zurlo} et~al.,}{{Zurlo}
  et~al.}{2020b}]{zurlo2020}
{Zurlo} A.,  et~al., 2020b, \mn@doi [\aap] {10.1051/0004-6361/201936891}, \href
  {https://ui.adsabs.harvard.edu/abs/2020A&A...633A.119Z} {633, A119}

\bibitem[\protect\citeauthoryear{{Zurlo} et~al.,}{{Zurlo}
  et~al.}{2021}]{zurlo2021}
{Zurlo} A.,  et~al., 2021, \mn@doi [\mnras] {10.1093/mnras/staa3674}, \href
  {https://ui.adsabs.harvard.edu/abs/2021MNRAS.501.2305Z} {501, 2305}

\makeatother
\end{thebibliography}




\appendix
\section{Effects of continuum over-subtraction}

\begin{figure*}
\begin{center}
\includegraphics[width=1\textwidth]{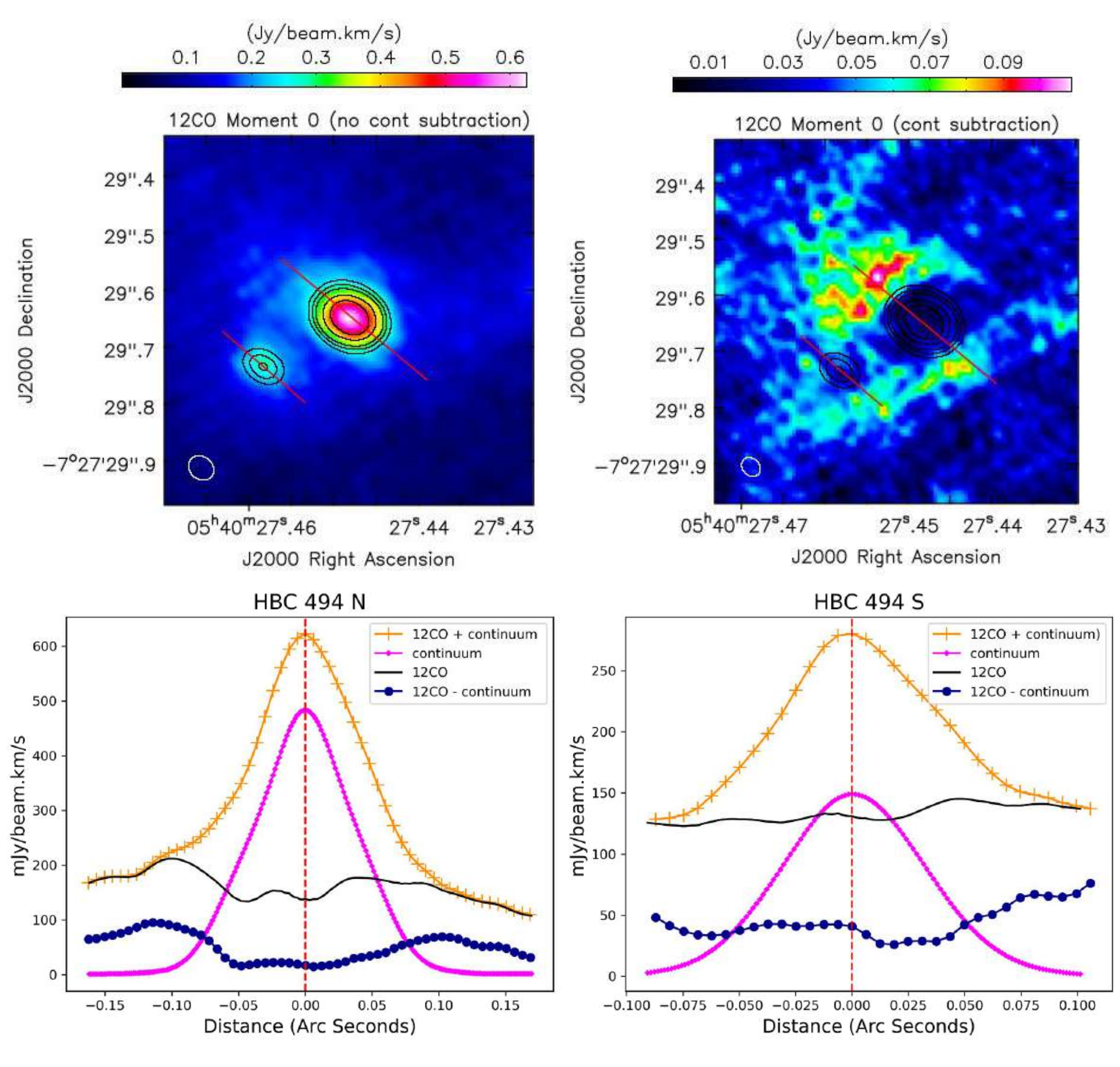}
\caption{The top panels show the $^{12}$CO moment 0 maps before (left) and after (right) continuum removal, along with their spatial profiles. The moment 0 contours represent values of 50, 75, 100, 150, and 200 times the rms ($\sim$0.06 mJy/beam.km/s). The red lines trace the axis used to create spatial profiles, which can be seen in the bottom panels. The bottom left panel shows the spatial profile of the HBC 494 N disk, while the bottom right panel shows that of the HBC 494 S disk. The vertical black dashed lines indicate the center of the disks. The orange crosses/lines represent spatial profiles before the continuum subtraction, created using the red line axis in the top panel, left. The magenta diamonds/lines represent the continuum spatial profile. The black lines represent the subtraction between the $^{12}$CO+continuum and the continuum spatial profiles for each disk. The dark blue filled circles/lines represent spatial profiles after the continuum subtraction, created using the red line axis in the top panel, right.}
\label{spatialprofiles}
\end{center}
\end{figure*}

To evaluate the effects of continuum over-subtraction, we generated spatial profiles for each disk using $^{12}$CO moment 0 maps at small scales before and after continuum subtraction. We created cuts that crossed the semi-major axis of each disk (see red lines in Figure \ref{spatialprofiles}, top panels) for this purpose. Figure \ref{spatialprofiles}, top-left panel displays moment 0 maps before continuum removal, while the top-right panel shows the same maps after this process. The comparison between these effects is better visualized in the bottom panels, where continuum emission (magenta diamonds/lines) has a similar shape as $^{12}$CO before continuum subtraction (orange crosses/lines), but shifted in flux, for both disks. The spatial profile subtractions are displayed by black lines. However, they display higher fluxes when compared to the spatial profiles produced after continuum subtraction (blue-filled circles/lines). Additionally, the continuum was overestimated by foreground optical depth molecular lines that absorbed continuum emission and then propagated to the \textit{uvcontsub} CASA task. The "holes" or cavities seen in the top-right panel could potentially be diminished by manually adding back some of the lost line signal, but this would only create a fake signal at the level of noise. Thus, optically thinner lines must be used to obtain the real signal of disks in moment 0 and 1 maps.

\section{$^{12}$CO, $^{13}$CO and C$^{18}$O channel maps with contours and the effect of cloud contamination}

\begin{figure*}
\begin{center}
\includegraphics[width=1\textwidth]{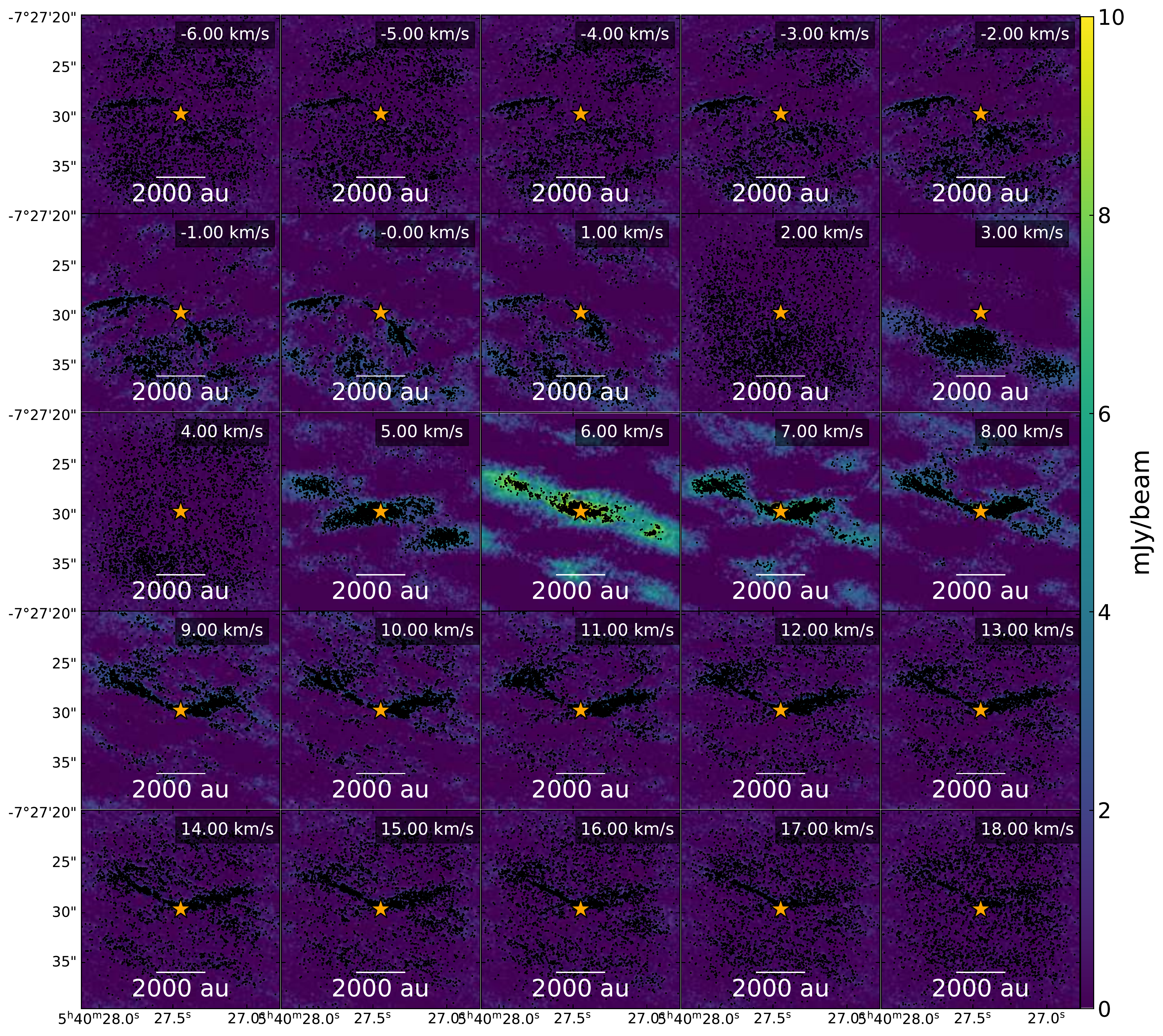}
\caption{$^{12}$CO channel maps of HBC 494 system with contours. The contours were displayed in black and they are equivalent to 3 times the rms of every channel. The rms values were calculated using all data in each channel, through the CASA task \textit{imstat}. The peak flux values have reached up to 19 times the rms values. The star in the center marks the position of the continuum disk around HBC 494 N.}
\label{12cochannelcontours}
\end{center}
\end{figure*}

\begin{figure*}
\begin{center}
\includegraphics[width=1\textwidth]{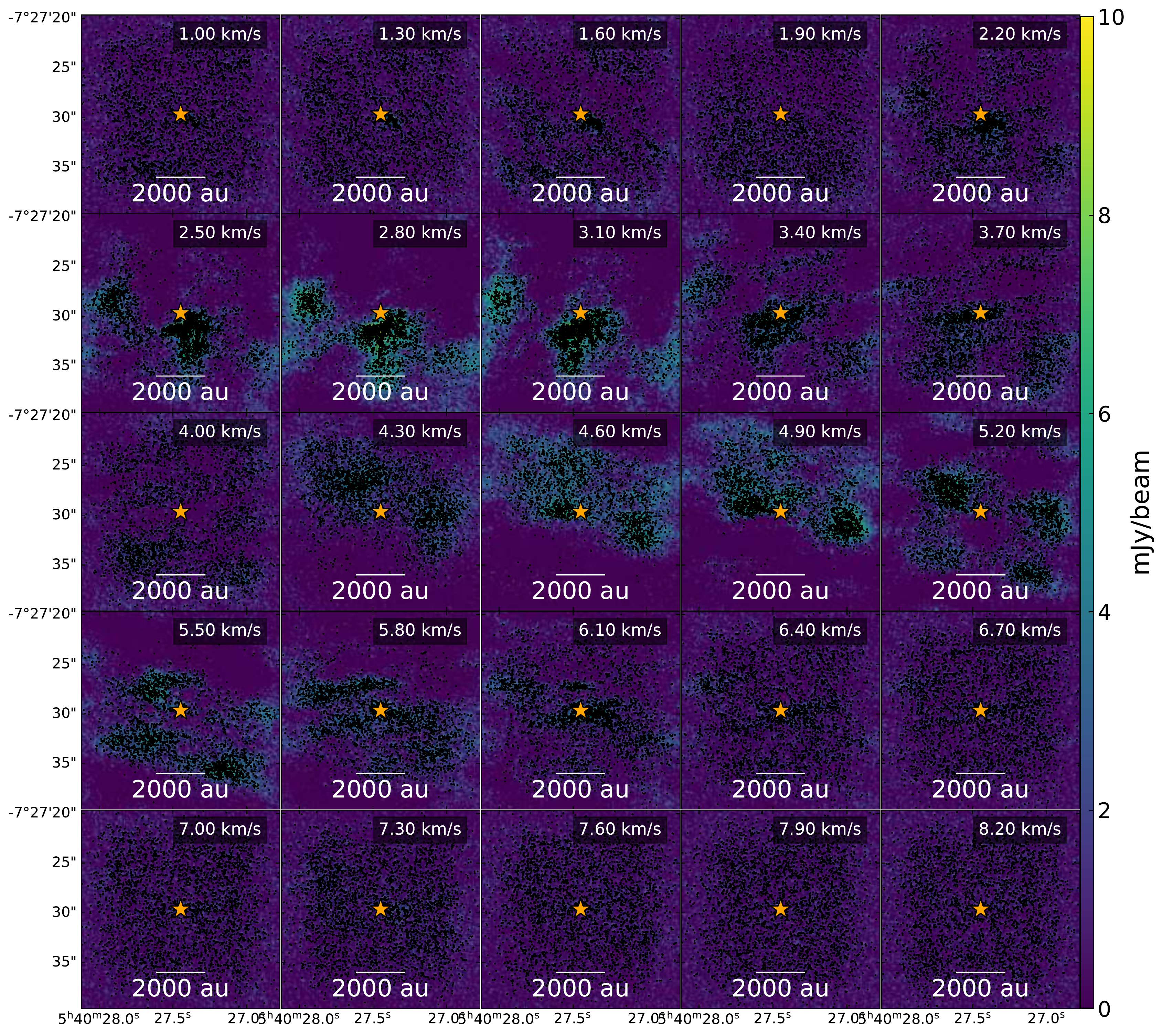}
\caption{$^{13}$CO channel maps of HBC 494 system with contours. The contours were displayed in black and they are equivalent to 3 times the rms of every channel. The rms values were calculated using all data in each channel, through the CASA task \textit{imstat}. The peak flux values have reached up to 11 times the rms values. The star in the center marks the position of the continuum disk around HBC 494 N.}
\label{13cochannelcontours}
\end{center}
\end{figure*}

This appendix presents the $^{12}$CO, $^{13}$CO, and C$^{18}$O channel maps with contour plots (see Figures \ref{12cochannelcontours} and \ref{13cochannelcontours}). To visualize the signals and corresponding flux values using a color bar, we also provide versions of the same channel maps without contours (see Figures \ref{12cochannel} and \ref{13cochannel}).

Figure \ref{12cochannelcontours} shows the 3-rms significant southern arcs (velocity channels from -6 km/s to 1 km/s) and northern arcs (velocity channels from 5 km/s to 18 km/s). However, cloud contamination affects channels between 2 km/s and 6 km/s. These channels were not used in the creation of large-scale moment maps. Channel 6 km/s shows the most evidence of cloud contamination, with contour plots showing innermost flux values near the disks (marked by the central orange star). It also shows some significant contours on both horizontal extremes, which can be tracers of the filtered signals using only the main array data. The data partially trace features that, due to their large extension, require a combination of the ALMA main and single dish arrays to recover extended flux emission.

In Figure \ref{13cochannelcontours}, the $^{13}$CO channel maps are heavily marked by 3-rms black contours, which show the significant gas signal on each channel. Specifically, channels from 2.20 km/s to 6.10 km/s indicate gas infalling clumps near where the disks are located, but also show random/scattered signals far from the disks due to cloud contamination.

\begin{figure*}
\begin{center}
\includegraphics[width=1\textwidth]{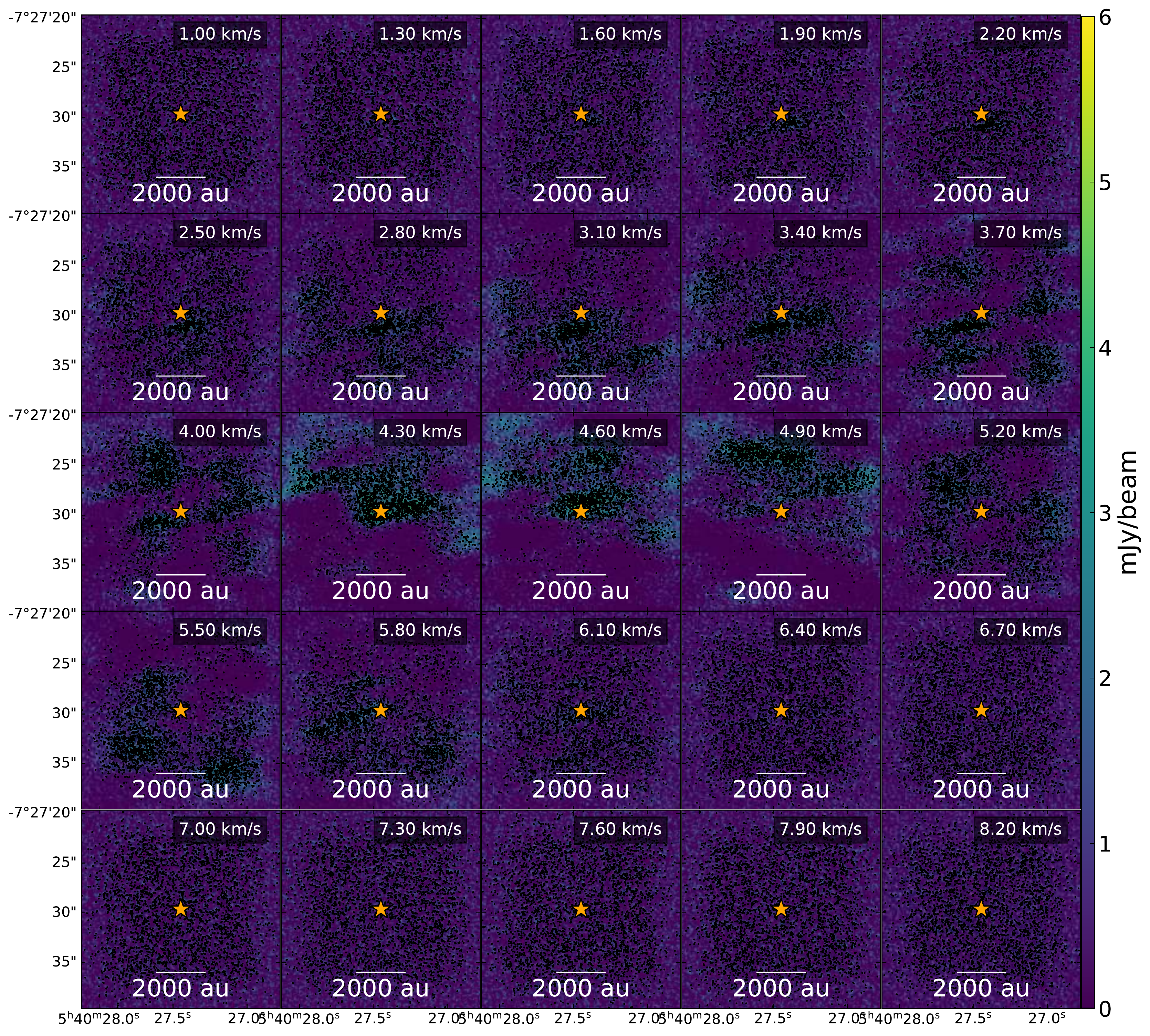}
\caption{C$^{18}$O channel maps of HBC 494 system with contours. The contours were displayed in black and they are equivalent to 3 times the rms of every channel. The rms values were calculated using all data in each channel, through the CASA task \textit{imstat}. The peak flux values have reached up to 13 times the rms values. The star in the center marks the position of the continuum disk around HBC 494 N.}
\label{c18ochannelcontours}
\end{center}
\end{figure*}

The C$^{18}$O channel maps represented in Figure \ref{c18ochannelcontours} are also heavily marked by 3-rms black contours. Channels from 2.80 km/s to 4.60 km/s indicate gas infalling clumps near where the disks are located, but also show random/scattered signals far from the disks due to cloud contamination. Other channels are affected by cloud contamination or lack a significant signal.

\section{Extra channel maps}

\begin{figure*}
\begin{center}
\includegraphics[width=1\textwidth]{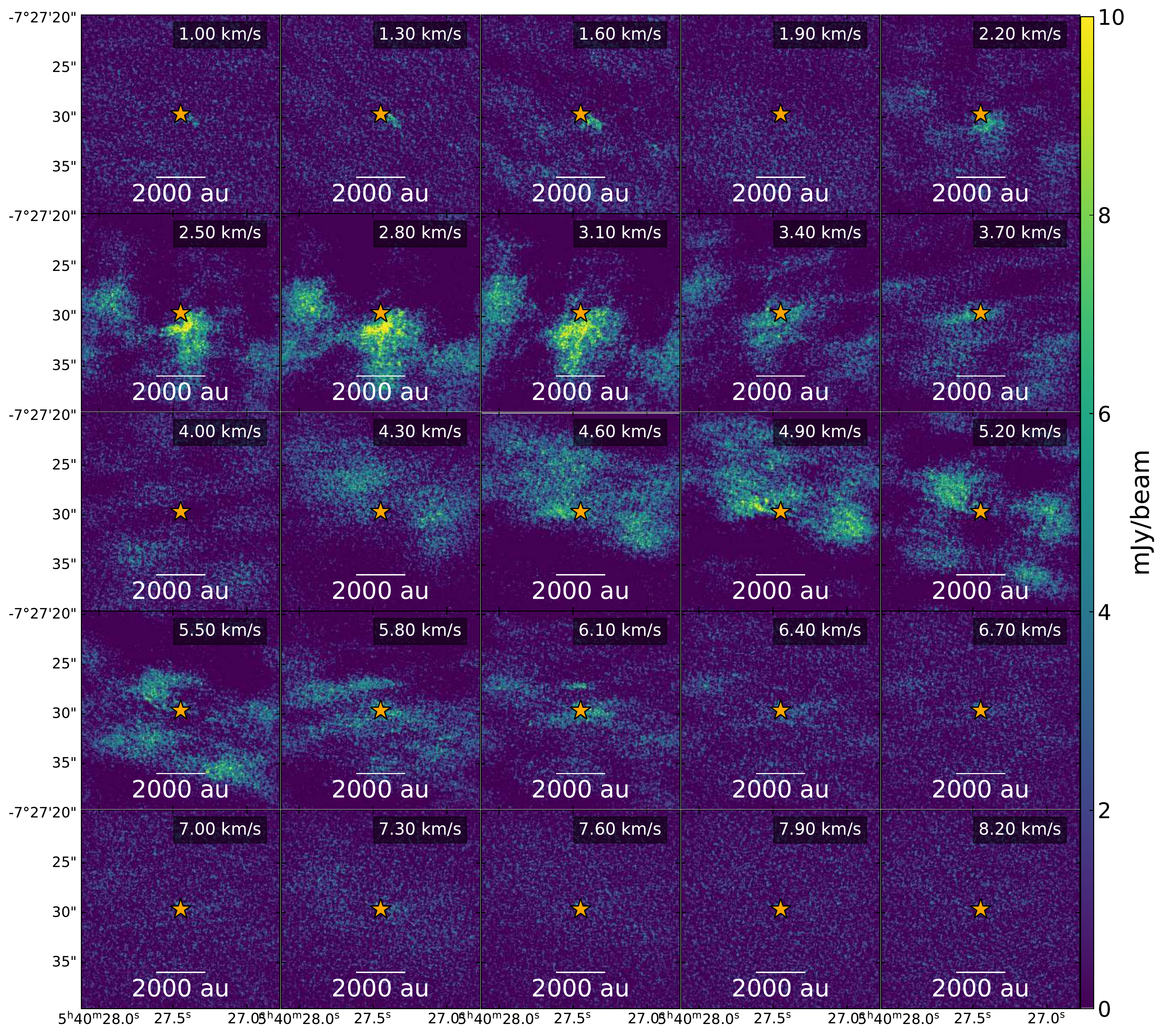}
\caption{$^{13}$CO channel maps of HBC 494 system. The star in the center marks the position of the continuum disk around HBC 494 N.}
\label{13cochannel}
\end{center}
\end{figure*}

\begin{figure*}
\begin{center}
\includegraphics[width=1\textwidth]{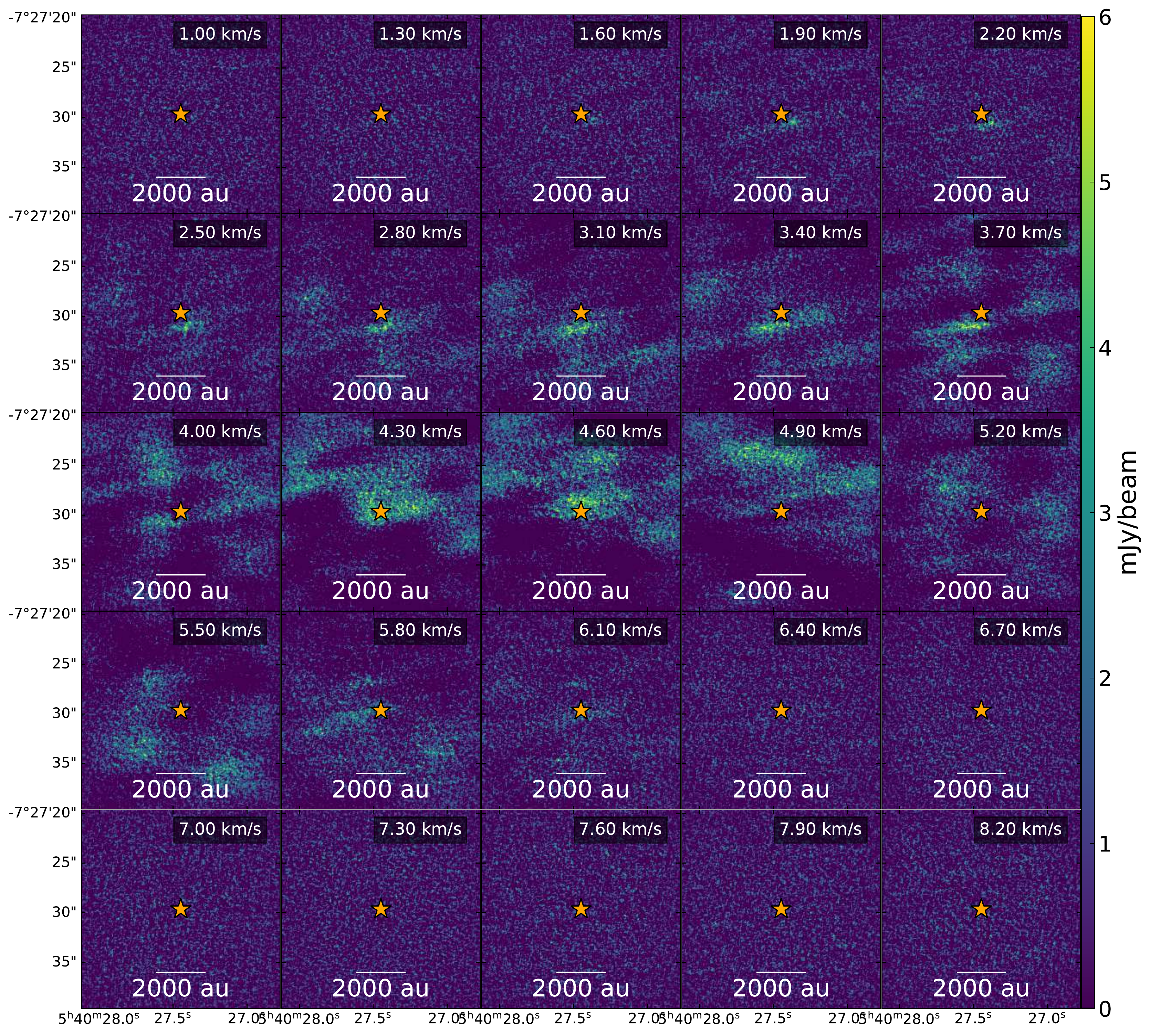}
\caption{C$^{18}$O channel maps of HBC 494 system.}
\label{c18ochannel}
\end{center}
\end{figure*}

\begin{figure*}
\begin{center}
\includegraphics[width=1\textwidth]{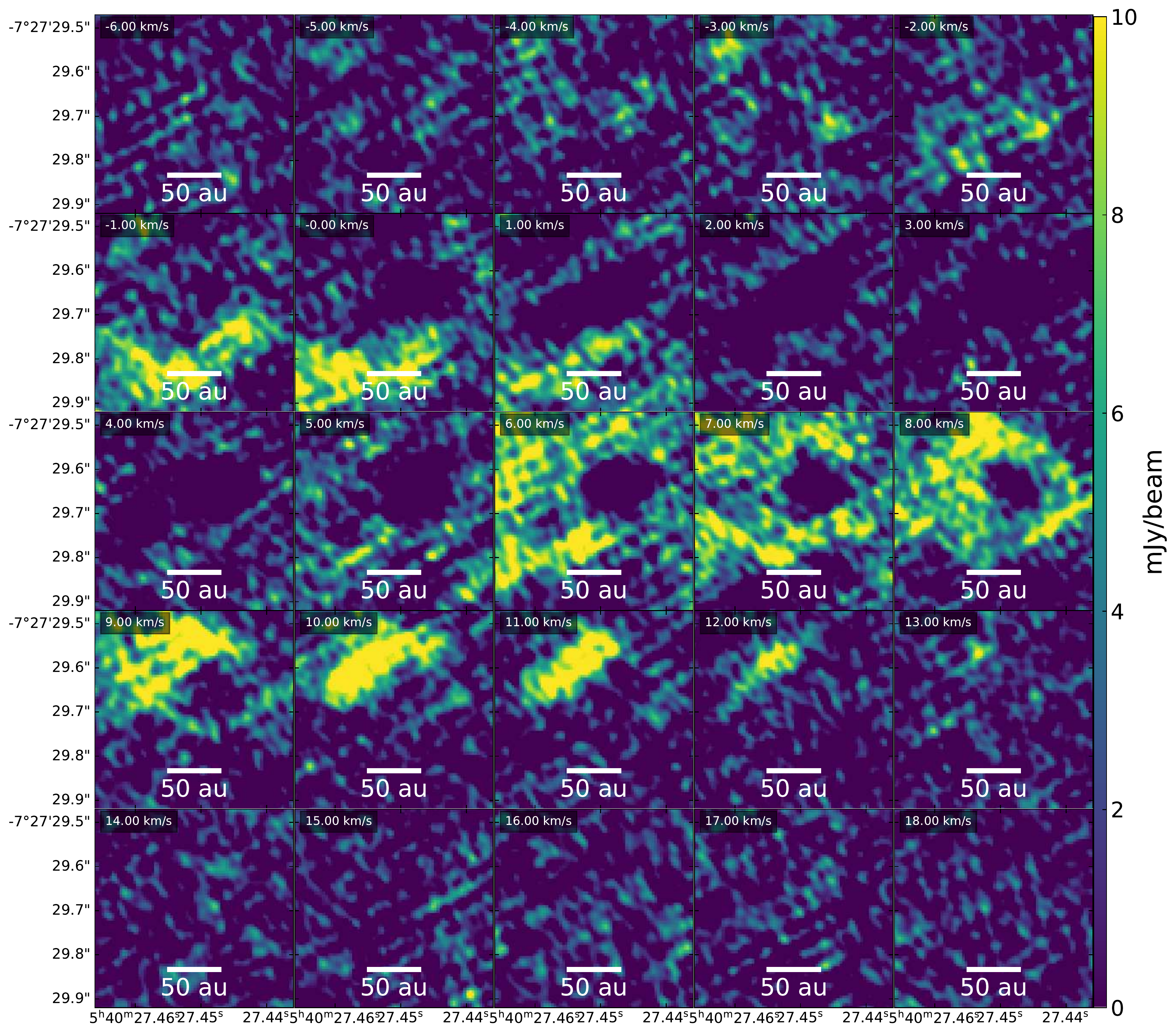}
\caption{$^{12}$CO channel maps of HBC 494 system, small-scale, after removing the continuum contribution.}
\label{12cochannel_nocont}
\end{center}
\end{figure*}

\begin{figure*}
\begin{center}
\includegraphics[width=1\textwidth]{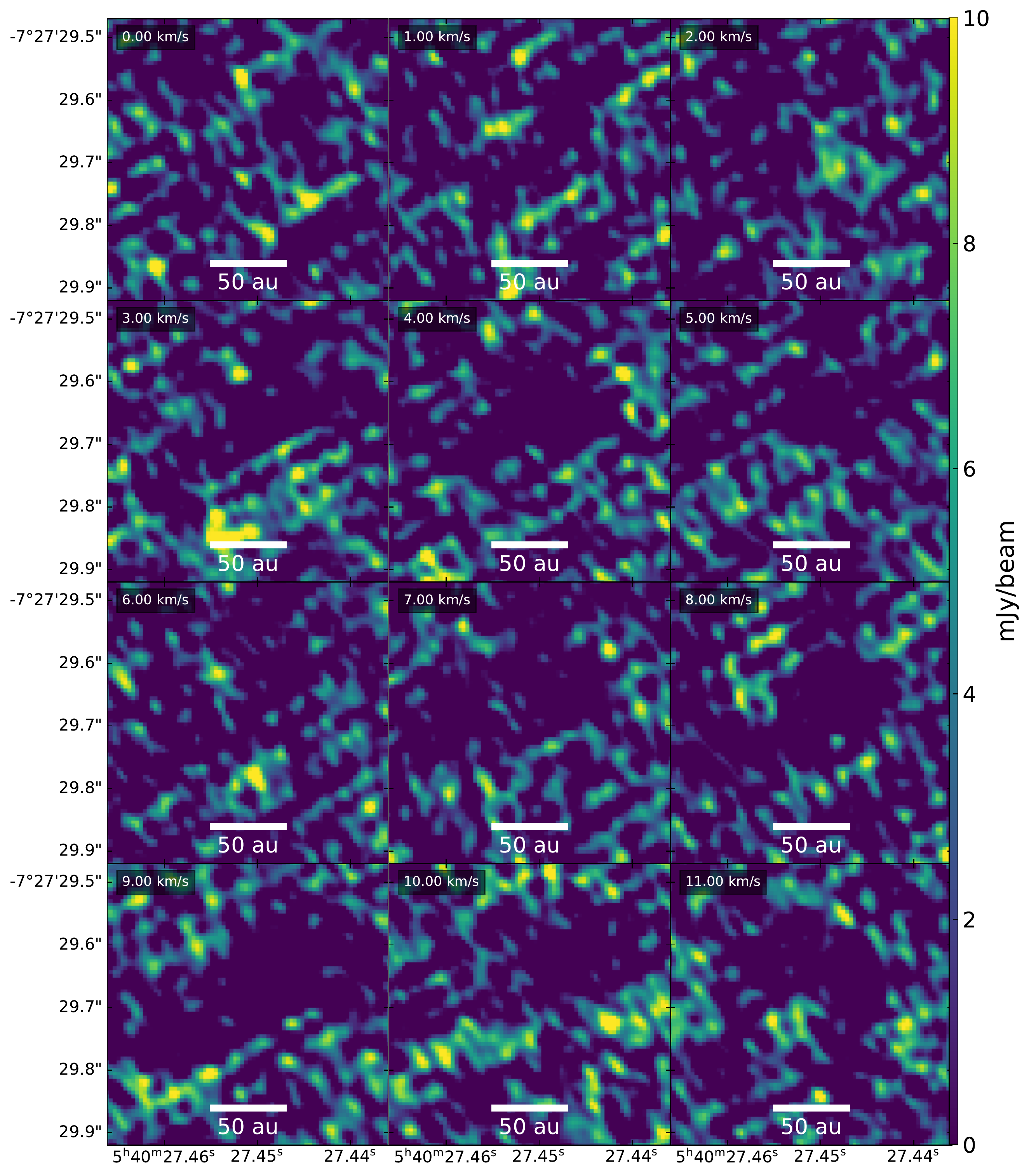}
\caption{$^{13}$CO channel maps of HBC 494 system, small-scale, after removing the continuum contribution.}
\label{13cochannel_nocont}
\end{center}
\end{figure*}

\begin{figure*}
\begin{center}
\includegraphics[width=1\textwidth]{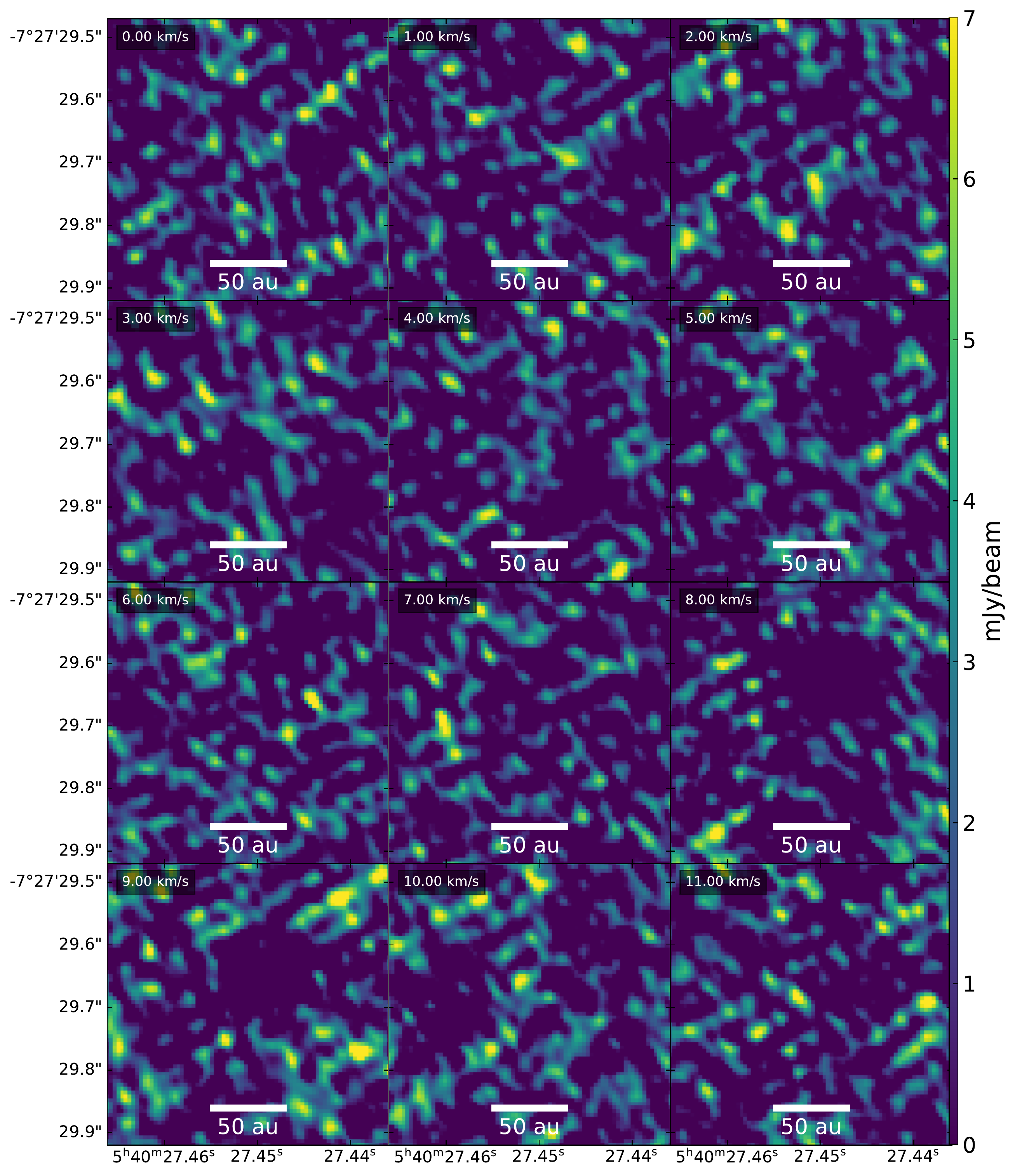}
\caption{C$^{18}$O channel maps of HBC 494 system, small-scale, after removing the continuum contribution.}
\label{c18ochannel_nocont}
\end{center}
\end{figure*}

The $^{13}$CO (large scale), C$^{18}$O (large scale), $^{12}$CO (small scale and with continuum subtracted), $^{13}$CO (small scale and with continuum subtracted), and C$^{18}$O (small scale and with continuum subtracted) channel maps are displayed in Figs. \ref{13cochannel}, \ref{c18ochannel}, \ref{12cochannel_nocont}, \ref{13cochannel_nocont}, and \ref{c18ochannel_nocont}, respectively.

\bsp	
\label{lastpage}
\end{document}